\documentclass{intech07a}
\usepackage{amsfonts}
\usepackage[latin1]{inputenc}
\usepackage{makeidx}

\input epsf.sty

\makeatletter \@addtoreset{equation}{section}
\renewcommand{\theequation}{\thesection.\arabic{equation}}
\input{tcilatex}

\booktitle{Graphene Simulation}

\headertext{Graphene and Cousin Systems }

\chaptertitle{Graphene and Cousin Systems}

\authors{L.B Drissi$^{1}$, E.H Saidi$^{1,2}$, M. Bousmina$^{1}$}
\affiliation{\makebox[3mm][l]{1}MAScIR-Inanotech, Institute for Nanomaterials and Nanotechnology, Rabat,\\
\makebox[3mm][l]{2}LPHE- Modelisation et Simulation, Facult\'{e} des
Sciences Rabat,}

\country{Morocco}

\begin{document}

\maketitle

\section{Introduction}

Besides its simple molecular structure, the magic of \emph{2D} graphene, a
sheet of carbon graphite, is essentially due to two fundamental electronic
properties: First for its peculiar band structure where valence and
conducting bands intersect at two points $K_{+}$ and $K_{-}$ of the
reciprocal space of the \emph{2D} honeycomb making of graphene a zero gap
semi-conductor. Second, for the ultra relativistic behavior of the charge
carriers near the Fermi level where the energy dispersion relation $%
E=E\left( p\right) $ behaves as a linear function in momenta; $E\left(
p\right) =v_{f}p+O\left( p^{2}\right) $. This typical property, which is
valid for particles with velocity comparable to the speed of light, was
completely unexpected in material science and was never suspected before
\emph{2004}; the year where a sheet of \emph{2D} graphene has been
experimentally isolated \textrm{\citep{A1,A2}}. From this viewpoint,
graphene is then a new material with exotic properties that could play a
basic role in the engineering of electronic devices with high performances;
it also offers a unique opportunity to explore the interface between
condensed interface between condensed matter physics and relativistic Dirac
theory where basic properties like chirality can be tested; and where some
specific features, such as numerical simulation methods, can be mapped to
\emph{4D} lattice gauge theory like lattice QCD \textrm{\citep{A3,A31,A32}}.
Although looking an unrealistic matter system, interest into the physical
properties of graphene has been manifested several decades ago. The first
model to analyze the band structure of graphite in absence of external
fields was developed by Wallace in \emph{1947 }\textrm{\citep{A4}}; see also
\textrm{\citep{A5}}. Since then, several theoretical studies have been
performed on graphene in the presence of a magnetic field \textrm{\citep{B1}-%
\citep{B5}}. The link between the electronic properties of graphene and $%
\left( 2+1\right) $-dimensional Dirac theory was also considered in many
occasions; in particular by Semenoff, Fradkin and Haldane during the \emph{80%
}-th of the last century \textrm{\citep{C1,C2,C5}}; see also \textrm{%
\citep{D1,D5} }and refs therein.\newline
In this book chapter, {we use the tight binding model as well as the }${SU}%
\left( {3}\right) ${\ hidden symmetry of \emph{2D} honeycomb to study some
physical aspects of \emph{2D} graphene with a special focus on the
electronic properties. We also develop new tools to study some of graphene's
cousin systems such as the \emph{1D}- poly-acetylene chain, cumulene,
poly-yne, Kekulé cycles, the \emph{3D} diamond and the \emph{4D}
hyperdiamond models. As another application of the physics in higher
dimension, we also develop the relation between the so called four
dimensional graphene first studied in \textrm{\citep{A3,E1,E2}}; and \emph{4D%
} lattice quantum chromodynamics (QCD) model considered recently in the
lattice quantum field theory (QFT) literature to deal with QCD numerical
simulations \textrm{\citep{F1,F2}}.}\newline
The presentation is as follows: \emph{In section 2}, we review the main
lines of the electronic properties of \emph{2D} graphene and show, amongst
others, that they are mainly captured by the $SU(3)$ symmetry of the \emph{2D%
} honeycomb. \emph{In section 3}, we study higher dimensional graphene type
systems by using the power of the hidden symmetries of the underlying
lattices. \emph{In section 4,} we give four examples of graphene's
derivatives namely the \emph{1D-} poly-acetylene chain, having a $SU(2)$
invariance, as well as Kekulé cycles thought of as a particular \emph{1D-}
system. We also study the \emph{3D} diamond model which exhibits a $SU\left(
4\right) $ symmetry; the corresponding \emph{2D} model, with $SU(3)$
invariance, is precisely the graphene considered in section 2. \emph{In
section 5}, we develop the four dimensional graphene model living on the
\emph{4D} hyperdiamond lattice with a SU$\left( 5\right) $ symmetry. \emph{%
In section 6}, we study an application of this method in the framework of
\emph{4D} lattice QCD. Last section is devoted to conclusion and comments.

\section{Two dimensional graphene}

First, we give a brief review on the tight binding modeling the physics of
\emph{2D} graphene; then we study its electronic properties by using hidden
symmetries. We show amongst others that the \emph{2D} honeycomb is precisely
the weight lattice of $SU\left( 3\right) $ \textrm{\citep{G1}}; and the two
Dirac points are given by the roots of $SU\left( 3\right) $. This study may
be also viewed as a first step towards building graphene type systems in
diverse dimensions.

\subsection{Tight binding model}

\begin{figure}[tbph]
\centering
\hspace{0cm} \includegraphics[width=6cm]{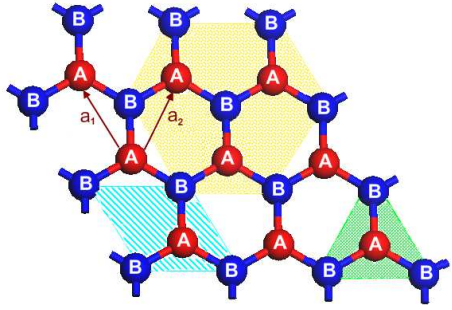}
\par
\caption{ Sublattices A and B of the honeycomb with unit cell
given by dashed area. A-type carbons are given by red balls and B-type atoms
by blue ones. Each carbon has three first nearest neighbors as shown by the
triangle; and six second nearest ones.}
\label{1}
\end{figure}

Graphene is a two dimensional matter system of carbon atoms in the $sp^{2}$\
hybridization forming a \emph{2D} honeycomb lattice. This is a planar system
made of two triangular sublattices $\mathcal{A}_{2}$ and $\mathcal{B}_{2}$;
and constitutes the building block of the layered \emph{3D} carbon graphite.
Since its experimental evidence in \emph{2004}, the study of the electronic
properties of graphene with and without external fields has been a big
subject of interest; some of its main physical aspects were reviewed in
\textrm{\citep{C5} }and refs therein. This big attention paid to the \emph{2D%
} graphene, its derivatives and its homologues is because they offer a real
alternative for silicon based technology\textrm{\ }and bring together issues
from condensed matter and high energy physics \textrm{\citep{G2}-\cite{G5} }%
allowing a better understanding of the electronic band structure as well as
their special properties\textrm{.} \newline
In this section, we focus on a less explored issue of \emph{2D} graphene by
studying the link between specific electronic properties and a class of
hidden symmetries of the \emph{2D} honeycomb. These symmetries allow to get
more insight into the transport property of the electronic wave modes and
may be used to approach the defects and the boundaries introduced in the
graphene monolayer \textrm{\citep{G6}}. The existence of these hidden
symmetries; in particular the remarkable hidden $SU\left( 3\right) $
invariance considered in this study, may be motivated from several views.
For instance from the structure of the first nearest carbon neighbors like
for the typical $\left\langle A_{0}\text{-}B_{1}\right\rangle $, $%
\left\langle A_{0}\text{-}B_{2}\right\rangle ,$ $\left\langle A_{0}\text{-}%
B_{3}\right\rangle $ as depicted in triangle of fig(\ref{1}).
These doublets $A_{0}$-$B_{1}$, $A_{0}$-$B_{2},$ $A_{0}$-$B_{3}$ are basic
patterns generating the three $SU\left( 2\right) $ symmetries contained in
the hidden $SU\left( 3\right) $ invariance of honeycomb. The $A$-$B$
patterns transform in the isospin $\frac{1}{2}$ representations of $SU\left(
2\right) $ and describe the electronic wave doublets $\phi _{{\scriptsize %
\pm }\frac{{\scriptsize 1}}{{\scriptsize 2}}}=\left[ a\left( \mathbf{r}%
\right) ,b\left( \mathbf{r}\right) \right] $ interpreted as
quasi-relativistic \emph{2D} spinors in the nearby of the Dirac points
\textrm{\citep{C5}}. The $SU\left( 3\right) $ hidden symmetry of honeycomb
is also encoded in the second nearest neighbors $\left\langle \left\langle
A_{0}\text{-}A_{i}\right\rangle \right\rangle $ and $\left\langle
\left\langle B_{0}\text{-}B_{i}\right\rangle \right\rangle ,$ $i=1,...,6$
which capture data on its adjoint representation where the six $\left\langle
\left\langle A_{0}\text{-}A_{i}\right\rangle \right\rangle $ (and similarly
for $\left\langle \left\langle B_{0}\text{-}B_{i}\right\rangle \right\rangle
$) are precisely associated with the six roots of $SU\left( 3\right) $
namely $\pm \mathbf{\alpha }_{1},$ $\pm \mathbf{\alpha }_{2},$ $\pm \mathbf{%
\alpha }_{3}$; see below. In addition to above mentioned properties, hidden
symmetries of graphene are also present in the framework of the tight
binding model with hamiltonian,
\vspace{-1mm}
\begin{equation}
\begin{tabular}{lll}
$H=$ & $-t\sum\limits_{\mathbf{r}_{i}}\sum\limits_{n=1}^{3}a_{\mathbf{r}%
_{i}}b_{\mathbf{r}_{i}{\scriptsize +}\mathbf{v}_{n}}^{\dagger }$ $-t^{\prime
}\sum\limits_{\mathbf{r}_{i}\mathbf{,r}_{j}}\left( a_{\mathbf{r}_{i}}a_{%
\mathbf{r}_{j}}^{\dagger }+b_{\mathbf{r}_{i}}b_{\mathbf{r}_{j}}^{\dagger
}\right) +hc$ & ,%
\end{tabular}
\label{HG}
\end{equation}%
\vspace{-1mm}
where $t\simeq 2.8eV$ is the hopping energy; and where the fermionic
creation and annihilation operators $a,$ $b,$ $a^{\dagger },$ $b^{\dagger }$
are respectively associated to the pi-electrons of each atom of the
sublattices $\mathcal{A}_{2}$ and $\mathcal{B}_{2}$. The three \emph{relative%
} vectors $\mathbf{v}_{1}$, $\mathbf{v}_{2}$, $\mathbf{v}_{3}$ define the
first nearest neighbors, see fig(\ref{2}) for illustration. These 2D vectors
are globally defined on the honeycomb and obey the remarkable constraint
equation
\vspace{-1mm}
\begin{equation}
\mathbf{v}_{1}+\mathbf{v}_{2}+\mathbf{v}_{3}=\mathbf{0,}  \label{de}
\end{equation}%
\vspace{-1mm}
which, a priori, encodes also information on the electronic properties of
graphene. Throughout this study, we show amongst others, that the three
above mentioned $SU\left( 2\right) $'s are intimately related with these $%
\mathbf{v}_{n}$'s which, as we will see, are nothing but the weight vectors $%
\mathbf{\lambda }_{n}$ of the $SU\left( 3\right) $ symmetry; i.e $\mathbf{v}%
_{n}=a\frac{\mathbf{\lambda }_{n}}{\left\Vert \mathbf{\lambda }%
_{n}\right\Vert }$. The wave functions $\phi _{\mathbf{\lambda }_{n}}\left(
\mathbf{r}\right) $ of the delocalized electrons are organized into a
complex $SU\left( 3\right) $ triplet of waves as given below%
\vspace{-1mm}
\begin{equation}
\begin{tabular}{lll}
$\left(
\begin{array}{c}
\left\vert \mathbf{\lambda }_{1}\right\rangle \\
\left\vert \mathbf{\lambda }_{2}\right\rangle \\
\left\vert \mathbf{\lambda }_{3}\right\rangle%
\end{array}%
\right) \equiv \underline{\mathbf{3}}$ & $,\qquad $ & $\mathbf{\lambda }_{1}+%
\mathbf{\lambda }_{2}+\mathbf{\lambda }_{3}=\mathbf{0.}$%
\end{tabular}%
\end{equation}%
\vspace{-1mm}
The symbol $\underline{\mathbf{3}}$ refers to the 3-dimensional
representation of $SU\left( 3\right) $; say with dominant weight $\mathbf{%
\lambda }_{1}$. We also show that the mapping of the condition $%
\sum_{n=1}^{3}\mathbf{\lambda }_{n}=\mathbf{0}$ to the momentum space can be
interpreted as a condition on the conservation of total momenta at each site
of honeycomb. This connection with $SU\left( 3\right) $ representations
opens a window for more insight into the study of the electronic
correlations in \emph{2D} graphene and its cousin systems by using
symmetries. \newline
The organization of this section is as follows: \emph{In subsection 2}, we
exhibit the $SU\left( 3\right) $ symmetry of graphene. We also give a field
theoretic interpretation of the geometric constraint equation $\mathbf{v}%
_{1}+\mathbf{v}_{2}+\mathbf{v}_{3}=\mathbf{0}$ both in real and reciprocal
honeycomb. We also use the simple roots and the fundamental weights of
hidden $SU\left( 3\right) $ symmetry to study aspects of the electronic
properties of \emph{2D} graphene. \emph{In subsection 3}, we develop the
relation between the energy dispersion relation $E\left( k_{x},k_{y}\right) $
and the hidden $SU\left( 3\right) $ symmetry. Comments regarding the link
between graphene bilayers\ and symmetries are also given.

\subsection{Symmetries and electronic properties}

\subsubsection{Hidden symmetries of graphene}

\begin{figure}[tbph]
\centering
\hspace{0cm} \includegraphics[width=8cm]{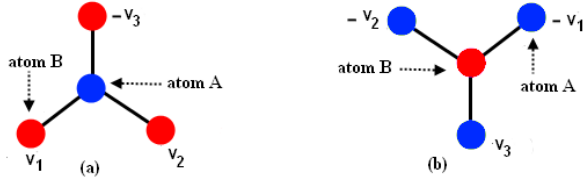}
\caption{ \ (a) Nearest neighbors of a A- type atom. (b)
Nearest neighbors of a B-type atom. These two configurations are precisely
the representations $ \mathbf{3}$ \ and $%
\mathbf{3}^{\ast }${\ of $ SU \left(
 3\right) $ .}}
\label{2}
\end{figure}

In dealing with pristine \emph{2D} graphene, one immediately notices the
existence of a hidden $SU\left( 3\right) $ group symmetry underlying the
crystallographic structure of the honeycomb lattice and governing the
hopping of the pi-electrons between the closed neighboring carbons. To
exhibit this hidden $SU\left( 3\right) $ symmetry, let us start by examining
some remarkable features on the graphene lattice and show how they are
closely related to \emph{SU}$\left( 3\right) $. Refereing to the two
sublattices of the graphene monolayer by the usual letters $\mathcal{A}_{2}$
and $\mathcal{B}_{2}$ generated by the vectors $\mathbf{a}_{1}=d(\sqrt{3}%
,0), $ $\mathbf{a}_{2}=\frac{d}{2}(-\sqrt{3},3)$; together with the three
relative $\mathbf{v}_{1}=\frac{d}{2}(\sqrt{3},1),$ $\mathbf{v}_{2}=\frac{d}{2%
}(-\sqrt{3},1),$ $\mathbf{v}_{3}=-\mathbf{v}_{1}-\mathbf{v}_{2}$ with
carbon-carbon distance $d\simeq 1.42$ $A^{°}$; and denoting by $\phi
_{A}\left( \mathbf{r}_{i}\right) $ and $\phi _{B}\left( \mathbf{r}%
_{j}\right) $ the wave functions of the corresponding pi-electrons, one
notes that the interactions between the first nearest atoms involve two
kinds of trivalent vertices capturing data on $SU\left( 3\right) $ symmetry,
see fig(\ref{2}) for illustration.
This hidden $SU\left( 3\right) $ invariance can be made more explicit by
remarking that the relative vectors $\mathbf{v}_{1},$ $\mathbf{v}_{2},%
\mathbf{v}_{3}$ describing the three first closed neighbors to a $\mathcal{A}
$- type carbon at site $\mathbf{r}_{i}$ of the honeycomb, together with
their opposites $-\mathbf{v}_{n}$ for $\mathcal{B}$-type carbons, are
precisely the weight vectors of the 3-dimensional representations of the $%
SU\left( 3\right) $ symmetry, $\mathbf{v}_{n}=d\sqrt{\frac{3}{2}}\mathbf{%
\lambda }_{n}$. For readers not familiar with representation group theory
terminology, we give here below as well in the beginning of section 3.a
summary on the $SU\left( N\right) $ symmetry.

\emph{some useful tools on SU}$\left( 3\right) $\newline
Roughly, the $SU\left( 3\right) $ symmetry is the simplest extension of the $%
SU\left( 2\right) $ symmetry group behind the spin of the electron. The
basic relation $\mathbf{v}_{1}+\mathbf{v}_{2}+\mathbf{v}_{3}=\mathbf{0}$ of
the honeycomb, which upon setting $\mathbf{v}_{n}=d\sqrt{\frac{3}{2}}\mathbf{%
\lambda }_{n}$, reads also as $\mathbf{\lambda }_{1}+\mathbf{\lambda }_{2}+%
\mathbf{\lambda }_{3}=\mathbf{0.}$ This constraint relation has an
interpretation in $SU\left( 3\right) $ representation theory; it should be
put in one to one correspondence with the well known $SU\left( 2\right) $
relation $(\frac{1}{2}-\frac{1}{2})=0$ of the spin $\frac{1}{2}$
representation;
\begin{equation}
\begin{tabular}{lll}
$\left(
\begin{array}{c}
\left\vert \frac{+1}{2}\right\rangle \\
\left\vert \frac{-1}{2}\right\rangle%
\end{array}%
\right) \equiv \underline{\mathbf{2}}$ & $,\qquad $ & $\frac{1}{2}-\frac{1}{2%
}=0\mathbf{,}$%
\end{tabular}%
\end{equation}%
see also eq(\ref{SU2}) for details. The basic properties of the $SU\left(
3\right) $ symmetry are encoded in the so called Cartan matrix $K_{ij}$ and
its inverse $K_{ij}^{-1}$ which read as%
\begin{equation}
\begin{tabular}{llll}
$K_{ij}=\left(
\begin{array}{cc}
2 & -1 \\
-1 & 2%
\end{array}%
\right) $ & ,$\qquad $ & $K_{ij}^{-1}=\left(
\begin{array}{cc}
\frac{2}{3} & \frac{1}{3} \\
\frac{1}{3} & \frac{2}{3}%
\end{array}%
\right) $ & .%
\end{tabular}
\label{K1}
\end{equation}%
These matrices can be also written as the intersection of 2D- vectors as $%
K_{ij}=\mathbf{\alpha }_{i}\cdot \mathbf{\alpha }_{j}$, $K_{ij}^{-1}=\mathbf{%
\omega }_{i}\cdot \mathbf{\omega }_{j}$ where $\mathbf{\alpha }_{1}$\ and $%
\mathbf{\alpha }_{2}$ are the two simple roots of $SU\left( 3\right) $ and
where $\mathbf{\omega }_{1}$\ and $\mathbf{\omega }_{2}$ are the
corresponding two fundamental weights which related to the simple roots by
the following duality relation%
\begin{equation}
\begin{tabular}{lll}
$\mathbf{\alpha }_{i}\cdot \mathbf{\omega }_{j}=\delta _{ij}$ & $,\qquad $ &
$\mathbf{\alpha }_{i}=K_{ij}\mathbf{\omega }_{j}.$%
\end{tabular}
\label{du}
\end{equation}%
Using these tools, the honeycomb relation $\mathbf{\lambda }_{1}+\mathbf{%
\lambda }_{2}+\mathbf{\lambda }_{3}=\mathbf{0}$ is naturally solved in terms
of the fundamental weights as follows%
\begin{equation}
\begin{tabular}{llll}
$\mathbf{\lambda }_{1}=\omega _{1},$ & $\mathbf{\lambda }_{2}=\omega
_{2}-\omega _{1},$ & $\mathbf{\lambda }_{3}=\omega _{2}$ & .%
\end{tabular}
\label{w}
\end{equation}%
We also have the following relations between the $\mathbf{a}_{i}$ vectors
and the $\mathbf{v}_{i}$ ones: $\mathbf{a}_{1}=\left( \mathbf{v}_{1}-\mathbf{%
v}_{2}\right) $, $\mathbf{a}_{2}=\mathbf{v}_{2}-\mathbf{v}_{3}$ and $\mathbf{%
a}_{3}=\mathbf{v}_{3}-\mathbf{v}_{1}$. Notice that the vectors $\pm \mathbf{a%
}_{1},$ $\pm \mathbf{a}_{2},$ $\pm \mathbf{a}_{3}$ are, up to the scale
factor $d\sqrt{\frac{3}{2}}$, precisely the six roots of the $SU\left(
3\right) $ symmetry
\begin{equation}
\begin{tabular}{lll}
$\mathbf{a}_{1}=d\sqrt{\frac{3}{2}}\mathbf{\alpha }_{1},$ & $\mathbf{a}_{2}=d%
\sqrt{\frac{3}{2}}\mathbf{\alpha }_{2},$ & $\mathbf{a}_{3}=-d\sqrt{\frac{3}{2%
}}\left( \mathbf{\alpha }_{1}+\mathbf{\alpha }_{2}\right) ,$%
\end{tabular}%
\end{equation}%
where we have also used the remarkable relation between roots and weights
that follow from eqs (\ref{K1}-\ref{du}).
\begin{equation}
\begin{tabular}{lll}
$\mathbf{\alpha }_{1}=2\omega _{1}-\omega _{2},$ & $\mathbf{\alpha }%
_{2}=2\omega _{2}-\omega _{1}$ &
\end{tabular}
\label{oa}
\end{equation}

\subsubsection{Electronic properties}

Quantum mechanically, there are two approaches to deal with the geometrical
constraint relation (\ref{de}). The first one is to work in real space and
think about it as the conservation law of total space-time probability
current densities at each site $\mathbf{r}_{i}$ of the honeycomb. The second
approach relies on moving to the reciprocal space where this constraint
relation and the induced electronic properties get a remarkable
interpretation in terms of $SU\left( 3\right) $ representations.

\textbf{1) conservation of total current density}\newline
In the real space, the way we interpret eq(\ref{de}) is in terms of the\
relation between the time variation of the probability density $\rho \left(
t,\mathbf{r}_{i}\right) =\left\vert \phi \left( t,\mathbf{r}_{i}\right)
\right\vert ^{2}$ of the electron at site $\mathbf{r}_{i}$ and the sum $%
\sum_{n=1}^{3}\mathbf{J}_{\mathbf{v}_{n}}\left( t,\mathbf{r}_{i}\right) =%
\mathbf{J}\left( t,\mathbf{r}_{i}\right) $\ of incoming and outgoing
probability current densities along the $\mathbf{v}_{n}$- directions. On one
hand, because of the \emph{equiprobability} in hopping from the carbon at $%
\mathbf{r}_{i}$ to each one of the three nearest carbons at $\mathbf{r}_{i}+%
\mathbf{v}_{n}$, the norm of the $\mathbf{J}_{\mathbf{v}_{n}}$- vector
current densities should be equal and so they should have the form
\begin{equation}
\begin{tabular}{llll}
$\mathbf{J}_{\mathbf{v}_{n}}\left( t,\mathbf{r}_{i}\right) =j\left( t,%
\mathbf{r}_{i}\right) \mathbf{e}_{n}$ & , & $n=1,2,3$ & .%
\end{tabular}%
\end{equation}%
These probability current densities together with the unit vectors $\mathbf{e%
}_{n}=\frac{\mathbf{v}_{n}}{d}$ pointing in the different $\mathbf{v}_{n}$-
direction; but have the same non zero norm: $\left\Vert \mathbf{J}_{\mathbf{v%
}_{1}}\right\Vert =\left\Vert \mathbf{J}_{\mathbf{v}_{2}}\right\Vert
=\left\Vert \mathbf{J}_{\mathbf{v}_{3}}\right\Vert =\left\vert j\right\vert $%
. Substituting in the above relation, the total probability current density $%
\mathbf{J}\left( t,\mathbf{r}_{i}\right) $ at the site $\mathbf{r}$ and time
$t$ takes then the factorized form
\begin{equation}
\mathbf{J}\left( t,\mathbf{r}\right) =\frac{j\left( t,r\right) }{d}\left(
\sum_{n}\mathbf{\lambda }_{n}\right) .  \label{tot}
\end{equation}%
On the other hand, by using the\ Schrodinger equation $i\hbar \frac{\partial
\phi }{\partial t}=\left( -\frac{\hbar ^{2}}{2m}\nabla ^{2}+V\right) \phi $
describing the interacting dynamics of the electronic wave at $\mathbf{r}$,
we have the usual conservation equation,%
\begin{equation}
\frac{\partial \rho \left( t,\mathbf{r}\right) }{\partial t}+\func{div}%
\mathbf{J}\left( t,\mathbf{r}\right) =0\text{ \ \ },
\end{equation}%
with probability density $\rho \left( t,\mathbf{r}\right) $ as before and $J=%
\frac{i\hbar }{2m}\left( \phi \nabla \phi ^{\ast }-\phi ^{\ast }\nabla \phi
\right) $ with m the mass of the electron and $\phi =\phi \left( t,\mathbf{r}%
\right) $ its wave. Moreover, assuming $\frac{\partial \rho }{\partial t}=0$
corresponding to stationary electronic waves $\phi \left( t,\mathbf{r}%
\right) =e^{i\omega t}\phi \left( \mathbf{r}\right) $, it follows that the
space divergence of the total current density vanishes identically; $\func{%
div}\mathbf{J}=0$. This constraint equation shows that generally $\mathbf{J}$
should be a curl vector; but physical consideration indicates that we must
have $\mathbf{J}\left( t,\mathbf{r}\right) =0$, in agreement with
Gauss-Stokes theorem $\int_{\mathcal{V}}\func{div}\mathbf{J}$ $d\mathcal{V}$
$\mathbf{=}$ $\int_{\partial \mathcal{V}}\mathbf{J.}d\mathbf{\sigma }$
leading to the same conclusion. Combining the property $\mathbf{J}\left( t,%
\mathbf{r}\right) =0$ with its factorized expression $\frac{j}{d}\left(
\sum_{n}\mathbf{v}_{n}\right) $ given by eq(\ref{tot}) together with $j\neq
0 $, we end with the constraint relation $\sum_{n}\mathbf{v}_{n}=0$.\

\textbf{2) conservation of total phase}\newline
In the dual space of the electronic wave of graphene, the constraint
relation (\ref{de}) may be interpreted in two different, but equivalent,
ways; first in terms of the conservation of the total relative phase $\Delta
\varphi _{{\scriptsize tot}}=\sum \mathbf{k.}\Delta \mathbf{r}$ of the
electronic waves induced by the hopping to the nearest neighbors. The second
way is in terms of the conservation of the total momenta at each site of the
honeycomb. \newline
Decomposing the wave function $\phi \left( \mathbf{r}\right) $, associated
with a A-type carbon at site $\mathbf{r}$, in Fourier modes as $\sum_{%
\mathbf{k}}e^{i2\pi \mathbf{k}\cdot \mathbf{r}}$ $\tilde{\phi}\left( \mathbf{%
k}\right) $; and similarly for the B-type neighboring ones $\phi \left(
\mathbf{r}+\mathbf{v}_{n}\right) =\sum_{\mathbf{k}}e^{i2\pi \mathbf{k}\cdot
\mathbf{r}}$ $\tilde{\phi}_{n}\left( \mathbf{k}\right) $ with $\mathbf{k=}%
\left( k_{x},k_{y}\right) $, we see that $\tilde{\phi}\left( k\right) $ and
the three $\tilde{\phi}_{n}\left( k\right) $ are related as
\begin{equation}
\begin{tabular}{lll}
$\tilde{\phi}_{n}\left( k\right) =e^{i2\pi \theta _{n}}\tilde{\phi}\left(
k\right) $ & , & $n=1,2,3$ \ ,%
\end{tabular}
\label{fn}
\end{equation}%
with relative phases $\theta _{n}=\mathbf{k}\cdot \mathbf{v}_{n}$. These
electronic waves have the same module, $\left\vert \tilde{\phi}_{n}\left(
k\right) \right\vert ^{2}=\left\vert \tilde{\phi}\left( k\right) \right\vert
^{2}$; but in general non zero phases; $\theta _{1}\neq \theta _{2}\neq
\theta _{3}$. This means that in the hop of an electron with momentum $%
\mathbf{p}=\hbar \mathbf{k}$ from a site $\mathbf{r}_{i}$ to the nearest one
at $\mathbf{r}_{i}+\mathbf{v}_{n}$, the electronic wave acquires an extra
phase of an amount $\theta _{n}$; but the probability density at each site
is invariant. Demanding the total relative phase to obey the natural
condition,
\begin{equation}
\theta _{1}+\theta _{2}+\theta _{3}=0\text{ \ },\text{ \ \ }\func{mod}\left(
2\pi \right) ,  \label{te}
\end{equation}%
one ends with the constraint eq(\ref{de}). Let us study two remarkable
consequences of this special conservation law on the $\theta _{n}$ phases by
help of the hidden $SU\left( 3\right) $ symmetry of graphene. Using eq(\ref%
{w}), which identifies the relatives $\mathbf{v}_{n}$\ vectors with the
weight vectors $\mathbf{\omega }_{n}$, as well as the duality relation $%
\mathbf{\alpha }_{i}\cdot \mathbf{\omega }_{j}=\delta _{ij}$ (\ref{du}), we
can invert the three equations $\theta _{n}=\mathbf{k}\cdot \mathbf{v}_{n}$
to get the momenta $\mathbf{p}_{n}\mathbf{=}\hbar \mathbf{k}_{n}$ of the
electronic waves along the $\mathbf{v}_{n}$-directions. For the two first $%
\theta _{n}$'s, that is $n=1,2$, the inverted relations are nicely obtained
by decomposing the 2D wave vector $\mathbf{k}$ along the $\mathbf{\alpha }%
_{1}$ and $\mathbf{\alpha }_{2}$ directions; that is $\mathbf{k}=k_{1}%
\mathbf{\alpha }_{1}+k_{2}\mathbf{\alpha }_{2}$; and end with the following
particular solution,%
\vspace{2mm}
\begin{equation}
\begin{tabular}{llll}
$\theta _{1}=k_{1}d,$ & $\theta _{2}=\left( k_{2}-k_{1}\right) d,$ & $\theta
_{3}=-k_{2}d$ & .%
\end{tabular}
\label{12}
\end{equation}
\vspace{2mm}

\subsection{Band structure}

We first study the case of graphene monolayer; then we extend the result to
the case of graphene bilayers by using the corresponding hidden symmetries.

\subsubsection{Graphene monolayer}

By considering a graphene sheet and restricting the tight binding
hamiltonian (\ref{HG}) to the first nearest neighbor interactions namely,%
\vspace{1mm}
\begin{equation}
H=-t\dsum\limits_{\mathbf{r}_{i}}\sum_{n=1}^{2}a_{\mathbf{r}_{i}}b_{\mathbf{r%
}_{i}+\mathbf{v}_{n}}^{\dagger }+hc,  \label{t}
\end{equation}%
\vspace{1mm}
we can determine the energy dispersion relation and the delocalized
electrons by using the $SU\left( 3\right) $ symmetry of the \emph{2D}
honeycomb. Indeed performing the Fourier transform of the various wave
functions, we end with the following expression of the hamiltonian in the
reciprocal space%
\vspace{1mm}
\begin{equation}
\begin{tabular}{lll}
$H=$ & $-t\sum\limits_{\mathbf{k}}\left( a_{\mathbf{k}}^{+},b_{\mathbf{k}%
}^{+}\right) \left(
\begin{array}{cc}
0 & \bar{\varepsilon}_{\mathbf{k}} \\
\varepsilon _{\mathbf{k}} & 0%
\end{array}%
\right) \left(
\begin{array}{c}
a_{\mathbf{k}} \\
b_{\mathbf{k}}%
\end{array}%
\right) $ & .%
\end{tabular}%
\end{equation}%
\vspace{1mm}
The diagonalization of this hamiltonian leads to the two eigenvalue $E_{\pm
}=\pm t\left\vert \varepsilon _{\mathbf{k}}\right\vert $ giving the energy
of the valence and conducting bands. In these relations, the complex number $%
\varepsilon _{\mathbf{k}}$ is an oscillating wave vector dependent function
given by $\varepsilon _{\mathbf{k}}=e^{idQ_{1}}+e^{idQ_{2}}+e^{-id\left(
Q_{1}+Q_{2}\right) }$ where we have set $Q_{l}=\mathbf{k}.\mathbf{\lambda }%
_{l}$. This relation, which is symmetric under permutation of the three $%
Q_{i}$, can be also rewritten by using the fundamental weights as follows,
\vspace{1mm}
\begin{equation}
\begin{tabular}{ll}
$\varepsilon _{\mathbf{k}}=\left( e^{id\left[ \mathbf{k}.\mathbf{\omega }_{1}%
\right] }+e^{id\left[ \mathbf{k}.\left( \mathbf{\omega }_{2}-\mathbf{\omega }%
_{1}\right) \right] }+e^{-id\left[ \mathbf{k}.\mathbf{\omega }_{2}\right]
}\right) $ & .%
\end{tabular}
\label{EK}
\end{equation}%
\vspace{1mm}
Up on expanding the wave vector as $\mathbf{k}=k_{1}\mathbf{\alpha }%
_{1}+k_{2}\mathbf{\alpha }_{2}$, this relation reads also as $\varepsilon _{%
\mathbf{k}}=e^{ik_{1}d}+e^{i\left( k_{2}-k_{1}\right) d}+e^{-ik_{2}d}$.
Notice that from (\ref{EK}), we learn that $\varepsilon _{\mathbf{k}}$ is
invariant under the translations $\mathbf{k}\rightarrow $ $\mathbf{k}+$ $%
\frac{2\pi }{d}\left( N_{1}\mathbf{\alpha }_{1}+N_{2}\mathbf{\alpha }%
_{2}\right) $ with $N_{1},$ $N_{2}$ arbitrary integers; thanks to the
duality relation $\mathbf{\alpha }_{i}.\mathbf{\omega }_{j}=\delta _{ij}$.
\newline
Notice also that near the origin $\mathbf{k}=\mathbf{0}$, we have $%
\varepsilon _{\mathbf{k}}=3+O\left( \mathbf{k}^{2}\right) $, in agreement
with non relativistic quantum mechanics. The three terms which are linear
terms in $\mathbf{k}$ cancel each others due to the $SU\left( 3\right) $
symmetry.
\begin{figure}[tbph]
\begin{center}
\hspace{0cm} \includegraphics[width=10cm]{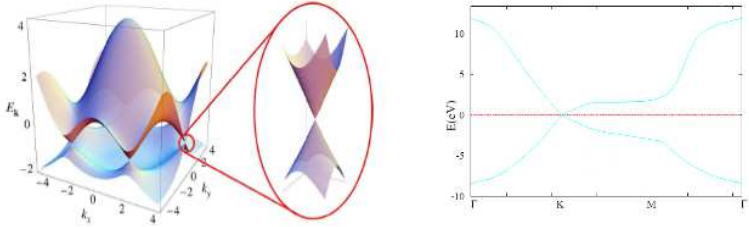}
\end{center}
\par
\vspace{-0.5 cm}
\caption{ On left the band structure of the graphene
monolayer where one recognizes the Dirac points. On right, it is shown the
relativistic behavior near a Dirac point where conducting and valence bands
touch. \ On right the band structure in GGA approximation using QE \ code.}
\label{BND}
\end{figure}
Notice moreover that the Hamiltonian (\ref{t}) has Dirac zeros located, up
to lattice translations, at the following wave vectors
\begin{equation}
\begin{tabular}{ll}
$\left( k_{1},k_{2}\right) =\left\{
\begin{array}{ccc}
\frac{2\pi }{3d}\left( 1,0\right) & , & -\frac{2\pi }{3d}\left( 1,0\right)
\\
\frac{2\pi }{3d}\left( 0,1\right) & , & -\frac{2\pi }{3d}\left( 0,1\right)
\\
\frac{2\pi }{3d}\left( 1,1\right) & , & -\frac{2\pi }{3d}\left( 1,1\right)%
\end{array}%
\right. $ &
\end{tabular}%
\end{equation}%
Notice that these six zero modes, which read also as
\begin{equation}
\begin{tabular}{lll}
$\mathbf{K}_{1}^{\pm }=\pm \frac{2\pi }{3d}\left( 2\mathbf{\omega }_{1}-%
\mathbf{\omega }_{2}\right) ,$ & $\mathbf{K}_{2}^{\pm }=\pm \frac{2\pi }{3d}%
\left( -\mathbf{\omega }_{1}+2\mathbf{\omega }_{2}\right) ,$ & $\mathbf{K}%
_{3}^{\pm }=\pm \frac{2\pi }{3d}\left( \mathbf{\omega }_{1}+\mathbf{\omega }%
_{2}\right) \mathbf{\ }$,%
\end{tabular}%
\end{equation}%
are not completely independent; some of them are related under lattice
translations. For instance, the three $\mathbf{K}_{i}^{+}$ are related to
each others as follows
\begin{equation}
\begin{tabular}{ll}
$\mathbf{K}_{1}^{+}+\frac{2\pi }{d}\mathbf{\omega }_{2}=\mathbf{K}_{2}^{+}+%
\frac{2\pi }{d}\mathbf{\omega }_{1}=\mathbf{K}_{3}^{+}$ & $.$%
\end{tabular}
\label{DP}
\end{equation}%
The same property is valid for the other three $\mathbf{K}_{i}^{-}$'s; so
one is left with the usual $\mathbf{K}_{\pm }$ Dirac zeros of the first
Brillouin zone,%
\begin{equation}
\mathbf{K}_{\pm }=\pm \frac{2\pi }{3d}\left( \mathbf{\omega }_{1}+\mathbf{%
\omega }_{2}\right) =\pm \frac{2\pi }{3d}\left( \mathbf{\alpha }_{1}+\mathbf{%
\alpha }_{2}\right) .  \label{2Z}
\end{equation}%
These two zeros are not related by lattice translations; but are related by
a $\mathbb{Z}_{2}$ symmetry mapping the fundamental weights and the simple
roots to their opposites. \newline
We end this section by noting that the group theoretical approach developed
in this study may be also used to deal with graphene multi-layers and cousin
systems. Below, we describe briefly the bilayers; the cousin systems are
studied in next sections.

\subsubsection{Bilayer graphene}

\begin{figure}[tbph]
\begin{center}
\hspace{0cm} \includegraphics[width=8cm]{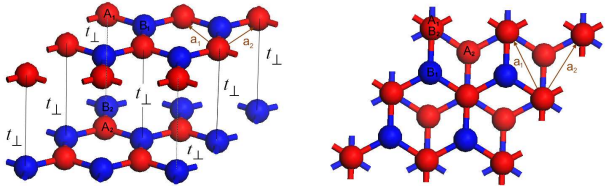}
\end{center}
\par
\caption{ On right bilayer graphene in the AB stacking
allotrope with hop energy t$_{\bot }$  between the layers.
On right bilayer graphene projection in the x-y plane. }
\label{X}
\end{figure}

Bilayer graphene was studied for the first time in\textrm{\ \cite{Bil1}}. It
was modeled as two coupled hexagonal lattices including inequivalent sites
in the two different layers that are ranged in the Bernal stacking (\emph{%
the stacking fashion of graphite where the upper layer has its B sublattice
on top of sublattice A of the underlying layer}) as showed in the figure (%
\ref{X}).
This leads to a break of the D$_{6h}$ Bravais symmetry of the lattice with \
respect to the c axis. Comparing bilayer graphene to monolayer one, we
notice that its unit cell contains four atoms. There exist other
arrangements such as the AA stacking, where the two lattices are directly
above each other and bonds form between the same sublattices. The AB
stacking arrangement was experimentally verified in epitaxial graphene by
Ohta et al. \textrm{\citep{Bil2}} . The tight-binding model describing
bilayer graphene is an extension of the one corresponding to the monolayer (%
\ref{HG}), by adding interlayer hopping elements $H=H_{1}+H_{2}+H_{\bot }$
where H$_{i}$ are as in (\ref{HG}) and where
\vspace{-1mm}
\begin{equation}
\begin{tabular}{lll}
$H_{\bot }=$ & $-t_{\bot }\sum\limits_{\mathbf{r}_{i}}\sum%
\limits_{n=1}^{3}a_{1}\left( \mathbf{r}_{i}\right) b_{2}^{\dagger }\left(
\mathbf{r}_{i}\right) +a_{2}\left( \mathbf{r}_{i}\right) b_{1}^{\dagger
}\left( \mathbf{r}_{i}\right) +hc$ & ,%
\end{tabular}%
\end{equation}%
\vspace{-1mm}
with $t_{\bot }$ is the hop energy of the pi-electrons between layers
calculated to be $t_{\bot }\sim \frac{t}{10}$ \textrm{\citep{Bil0}}. From
the view of hidden symmetries, the bilayer graphene has a symmetry type $%
SU\left( 3\right) \times SU\left( 2\right) \times SU\left( 3\right) $; each $%
SU\left( 3\right) $ factor is associated with a graphene sheet; while the $%
SU\left( 2\right) $ corresponds to the transitions between the two layers
and is associated with propagation along the z-direction of the \emph{3D}%
-space. \newline
Applying Fourier transform, the above hamiltonian can be rewritten in the
following form:%
\vspace{-1mm}
\begin{equation}
\begin{tabular}{lll}
$H=$ & $-t\sum\limits_{\mathbf{k}}\left( a_{1\mathbf{k}}^{+},b_{1\mathbf{k}%
}^{+},a_{2\mathbf{k}}^{+},b_{2\mathbf{k}}^{+}\right) \left(
\begin{array}{cccc}
0 & \varepsilon _{k} & 0 & \frac{t_{\bot }}{t} \\
\varepsilon _{k}^{\ast } & 0 & \frac{t_{\bot }}{t} & 0 \\
0 & \frac{t_{\bot }}{t} & 0 & \varepsilon _{k} \\
\frac{t_{\bot }}{t} & 0 & \varepsilon _{k}^{\ast } & 0%
\end{array}%
\right) \left(
\begin{array}{c}
a_{1\mathbf{k}} \\
b_{1\mathbf{k}} \\
a_{2\mathbf{k}} \\
b_{2\mathbf{k}}%
\end{array}%
\right) $ & ,%
\end{tabular}%
\end{equation}%
\vspace{-1mm}
with $\varepsilon _{\mathbf{k}}$ is as in eq(\ref{EK}). The diagonalization
of this hamiltonian leads to the following energy dispersion relations,%
\vspace{-1mm}
\begin{equation}
\begin{tabular}{lll}
$E_{\mathbf{k}}^{\pm }=\pm \frac{1}{t}\sqrt{\left( t_{\bot }-t\varepsilon _{%
\mathbf{k}}^{\ast }\right) \left( t_{\bot }-t\varepsilon _{\mathbf{k}%
}\right) },$ & $E_{k}^{\pm \prime }=\pm \frac{1}{t}\sqrt{\left( t_{\bot
}+t\varepsilon _{\mathbf{k}}^{\ast }\right) \left( t_{\bot }+t\varepsilon _{%
\mathbf{k}}\right) }$ & ,%
\end{tabular}%
\end{equation}%
\vspace{-1mm}
The corresponding band structure has two additional bands, $\pi $ and $\pi
^{\ast }$ states having lower energy bands, that is consequence of the
number of atoms per unit cell. Neutral bilayer graphene is gapless \textrm{%
\cite{Bil1}} and exhibits a variety of second-order effects.
\begin{figure}[tbph]
\begin{center}
\hspace{0cm} \includegraphics[width=6cm]{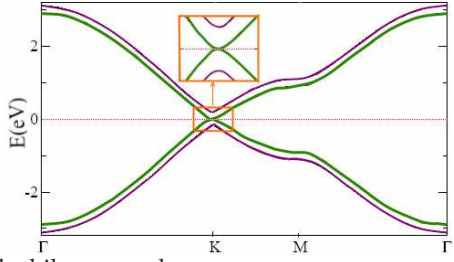}
\end{center}
\par
\vspace{-0.5 cm}
\caption{ Band structure of the bilayer graphene}
\label{2BB}
\end{figure}
The studies on bilayer graphene show that it has many common physical
properties with the monolayer, such as the exceptionally high electron
mobility and high mechanical stability \textrm{\citep{Bil2}-\cite{Bil3}}.\
The synthesis of bilayer graphene thin films was realized by deposition on a
silicon carbide (\emph{SiC}) substrate \textrm{\citep{Bil5}}. The
measurements of their electronic band structure, using angle-resolved
photo-emission spectroscopy (\emph{ARPES}), suggest the control of the gap
at the K Point by applying Coulomb potential between the two layers. This
tuning of the band gap changed the biased bilayer from a conductor to a
semiconductor.

\section{Higher dimensional graphene systems}

Motivated by the connection between \emph{2D} graphene and $SU\left(
3\right) $ symmetry, we study in this section the extension of the physics
of \emph{2D} graphene in diverse dimensions; that is \emph{1D}, \emph{2D},
\emph{3D}, \emph{4D}, and so on; the \emph{2D} case is obviously given by
\emph{2D} graphene and its multi-layers considered in previous section. The
precited dimensions are not all of them realizable in condensed matter
physics; but their understanding may help to get more insight on the
specific properties of \emph{2D} graphene since the $SU\left( 3\right) $ is
the second element of the $SU\left( N\right) $ symmetries series. \newline
First we develop our proposal regarding higher dimensional graphene systems
that are based on $SU\left( N\right) $ symmetry including the particular
\emph{1D} poly-acetylene chain which corresponds to $SU\left( 2\right) $
symmetry. Then, we compute the energy dispersion relation of these kinds of
lattice quantum field theory (QFT). Explicit examples of such lattice
fermionic models will be studied in the next sections.

\subsection{The $SU\left( N\right) $ model}

Higher dimensional graphene systems are abstract extensions of \emph{2D}
graphene; the analogue of the \emph{2D} honeycomb is given by a real
N-dimensional lattice $\mathcal{L}_{su\left( N\right) }$. The quantum
hamiltonian describing these systems is a generalization of (\ref{HG}) and
reads as follows,
\begin{equation}
\begin{tabular}{lll}
$H_{N}=$ & $-t\sum\limits_{\mathbf{r}_{i}}\left( \sum\limits_{n=1}^{N}a_{%
\mathbf{r}_{i}}b_{\mathbf{r}_{i}{\scriptsize +}\mathbf{v}_{n}}^{\dagger
}\right) +hc-t^{\prime }\sum\limits_{\mathbf{r}_{i}}\sum\limits_{n<m=1}^{N}%
\left( a_{\mathbf{r}_{i}}a_{\mathbf{r}_{i}+\mathbf{V}_{nm}}^{\dagger }+b_{%
\mathbf{r}_{i}}b_{\mathbf{r}_{i}+\mathbf{V}_{nm}}^{\dagger }\right) ,$ &
\end{tabular}
\label{HN}
\end{equation}%
where $a_{\mathbf{r}_{i}}$, $b_{\mathbf{r}_{i}{\scriptsize +}\mathbf{v}_{n}}$%
, $a_{\mathbf{r}_{i}}^{\dagger }$, $b_{\mathbf{r}_{i}{\scriptsize +}\mathbf{v%
}_{n}}^{\dagger }$ are fermionic annihilation and creation operators living
on $\mathcal{L}_{su\left( N\right) }$. Moreover the vectors $\mathbf{v}%
_{1},...,\mathbf{v}_{N}$ are, up to a global scale factor, the fundamental
weights of the N-dimensional representation of the $SU\left( N\right) $
symmetry constrained by the typical property
\begin{equation}
\mathbf{v}_{1}+...+\mathbf{v}_{N}=0.  \label{v}
\end{equation}%
The vectors $\mathbf{V}_{nm}=\left( \mathbf{v}_{n}-\mathbf{v}_{m}\right) $
are, up to a scale factor, precisely the $N\left( N-1\right) $ roots of $%
SU\left( N\right) $; they obey as well the group property $\sum \mathbf{V}%
_{nm}=0$. \newline
These particular features of $H_{N}$ let understand that its physical
properties are expected to be completely encoded by the hidden $SU\left(
N\right) $ symmetry of the model. Below, we show that this is indeed the
case; but for simplicity we will focus on the first term of $H_{N}$; i.e
working in the limit $t^{\prime }\rightarrow 0$.

\subsubsection{Useful tools on $SU\left( N\right) $ symmetry}

Since one of our objectives in this paper is to use the $SU\left( N\right) $
symmetry of the crystals to study higher dimensional graphene systems; and
seen that readers might not be familiar with these tools; we propose to give
in this subsection some basic tools on $SU\left( N\right) $ by using
explicit examples.

\emph{a)} \emph{cases} $SU\left( 2\right) $ and $SU\left( 3\right) $\newline
The $SU\left( 2\right) $ symmetry is very familiar in quantum mechanics; it
is the symmetry that describes the spin of the electrons and the quantum
angular momentum states. \newline
Roughly speaking, the $SU\left( 2\right) $ symmetry is a 3-dimensional space
generated by three matrices which can be thought of as the usual traceless
Pauli matrices%
\begin{equation}
\begin{tabular}{llll}
$\sigma ^{0}=\left(
\begin{array}{cc}
\frac{1}{2} & 0 \\
0 & -\frac{1}{2}%
\end{array}%
\right) ,$ & $\sigma ^{-}=\left(
\begin{array}{cc}
0 & 0 \\
1 & 0%
\end{array}%
\right) ,$ & $\sigma ^{+}=\left(
\begin{array}{cc}
0 & 1 \\
0 & 0%
\end{array}%
\right) $ & ,%
\end{tabular}
\label{SU2}
\end{equation}%
involving one diagonal matrix $\sigma ^{0},$ giving the charge operator, and
two nilpotent matrices $\sigma ^{\pm }$ interpreted as the step operators or
equivalently the creation and annihilation operators in the language of
quantum mechanics. These three matrices obey commutation relations $\left[
\sigma ^{0},\sigma ^{\pm }\right] =\pm 2\sigma ^{\pm }$ that define the $%
su\left( 2\right) $ algebra. Observe also the traceless property of the
charge operator $Tr\sigma ^{0}=\frac{1}{2}-\frac{1}{2}=0$, which should be
related to the constraint relation (\ref{v}) with $N=2$.\newline
The $SU\left( 3\right) $ symmetry group is 8-dimensional space generated by
\emph{8} matrices which can be denoted as
\begin{equation}
\begin{tabular}{lll}
$h_{1},h_{2},$ & $e^{\pm \alpha _{1}},e^{\pm \alpha _{2}},e^{\pm \left(
\alpha _{1}+\alpha _{2}\right) }$ & ,%
\end{tabular}%
\end{equation}%
with $h_{1},$ $h_{2}$ two diagonal matrices defining the charge operators
and six step operators $e^{\pm \alpha _{1}},$ $e^{\pm \alpha _{2}},$ $e^{\pm
\alpha _{3}}$ playing the role of creation and annihilation operators. The $%
e^{\pm \alpha _{i}}$'s are nilpotent and are related as $\left( e^{-\alpha
_{i}}\right) ^{+}=e^{+\alpha _{i}}$. An example of these matrices is given
by the following $3\times 3$ matrices%
\begin{equation}
\begin{tabular}{lll}
\multicolumn{3}{l}{$\ \ \ \ \ \ \ \ \ \ \ {\small h}_{1}{\small =}\left(
\begin{array}{ccc}
{\small \mu }_{1} & {\small 0} & {\small 0} \\
{\small 0} & {\small \mu }_{2} & {\small 0} \\
{\small 0} & {\small 0} & {\small \mu }_{3}%
\end{array}%
\right) ,$ $\ \ {\small h}_{2}{\small =}\left(
\begin{array}{ccc}
{\small \mu }_{1}^{\prime } & {\small 0} & {\small 0} \\
{\small 0} & {\small \mu }_{2}^{\prime } & {\small 0} \\
{\small 0} & {\small 0} & {\small \mu }_{3}^{\prime }%
\end{array}%
\right) ,$} \\
&  &  \\
\multicolumn{3}{l}{${\small e}^{+\alpha _{1}}{\small =}\left(
\begin{array}{ccc}
{\small 0} & {\small 1} & {\small 0} \\
{\small 0} & {\small 0} & {\small 0} \\
{\small 0} & {\small 0} & {\small 0}%
\end{array}%
\right) ,\ \ \ {\small e}^{+\alpha _{2}}{\small =}\left(
\begin{array}{ccc}
{\small 0} & {\small 0} & {\small 0} \\
{\small 0} & {\small 0} & {\small 1} \\
{\small 0} & {\small 0} & {\small 0}%
\end{array}%
\right) ,\ \ \ {\small e}^{+\alpha _{3}}{\small =}\left(
\begin{array}{ccc}
{\small 0} & {\small 0} & {\small 1} \\
{\small 0} & {\small 0} & {\small 0} \\
{\small 0} & {\small 0} & {\small 0}%
\end{array}%
\right) $} \\
&  &
\end{tabular}%
\end{equation}%
with the traceless property of the charge operators which reads as follows%
\begin{equation}
\begin{tabular}{llll}
$\mathbf{\lambda }_{1}+\mathbf{\lambda }_{2}+\mathbf{\lambda }_{3}=0$ & , & $%
\mathbf{\lambda }_{i}=\left(
\begin{array}{c}
\mu _{i} \\
\mu _{i}^{\prime }%
\end{array}%
\right) $ & ,%
\end{tabular}%
\end{equation}%
and which should be compared with the case $N=3$ in (\ref{v}).\newline
The vectors $\mathbf{\alpha }_{1}$ and $\mathbf{\alpha }_{2}$ are the simple
roots encountered in the previous section; their scalar product $\mathbf{%
\alpha }_{i}.\mathbf{\alpha }_{j}$ gives precisely the Cartan matrix $%
\mathbf{K}_{ij}$ of eq(\ref{K1}).

\emph{b)} \emph{case }$SU\left( N\right) $\newline
In the general case $N\geq 2$, the corresponding $SU\left( N\right) $
symmetry is $\left( N^{2}-1\right) $-dimensional space generated by $\left(
N^{2}-1\right) $ matrices; $N-1$ of them are diagonal%
\begin{equation}
\begin{tabular}{llll}
$h_{1},$ & $...$ & $h_{N-1},$ &
\end{tabular}%
\end{equation}%
and are interpreted as the charge operators; and $N\left( N-1\right) $ step
operators giving the creation and annihilation operators $e^{+\alpha },$ $%
e^{-\alpha }$ with $\mathbf{\alpha }$ standing for generic roots containing
the two following:\newline
(\textbf{a}) \emph{N-1} simple ones namely $\mathbf{\alpha }_{1}\mathbf{%
,\alpha }_{2}\mathbf{,...,\alpha }_{N-1}$ (together with their opposites)
whose scalar products $\mathbf{\alpha }_{i}.\mathbf{\alpha }_{j}$ give
precisely the following $\left( N-1\right) \times \left( N-1\right) $ Cartan
matrix%
\begin{equation}
\mathbf{K}=\left(
\begin{array}{cccccc}
2 & -1 & 0 & \cdots & 0 & 0 \\
-1 & 2 & -1 &  & 0 & 0 \\
0 & -1 & 2 &  & 0 & 0 \\
\vdots &  &  & \ddots &  & \vdots \\
0 & 0 & 0 &  & 2 & -1 \\
0 & 0 & 0 & \cdots & -1 & 2%
\end{array}%
\right) ,
\end{equation}%
\textbf{(b)} non simple roots given by linear (positive and negative)
combinations of the simple ones; these roots are given by $\pm \mathbf{\beta
}_{ij}=\pm \left( \mathbf{\alpha }_{i}+...+\mathbf{\alpha }_{j}\right) $
with $1\leq i<j\leq N-1.$\newline
Notice that the above Cartan matrix $\mathbf{K}$ and its inverse
\begin{equation}
\mathbf{K}^{-1}=\left(
\begin{array}{cccccc}
\frac{N}{N+1} & \frac{N-1}{N+1} & \frac{N-2}{N+1} & \cdots & \frac{2}{N+1} &
\frac{1}{N+1} \\
\frac{N-1}{N+1} & \frac{2\left( N-1\right) }{N+1} & \frac{2\left( N-2\right)
}{N+1} & \cdots & \frac{4}{N+1} & \frac{2}{N+1} \\
\frac{N-2}{N+1} & \frac{2\left( N-2\right) }{N+1} & \frac{3\left( N-2\right)
}{N+1} & \cdots & \frac{6}{N+1} & \frac{3}{N+1} \\
\vdots & \vdots & \vdots & \ddots & \vdots & \vdots \\
\frac{2}{N+1} & \frac{4}{N+1} & \frac{6}{N+1} & \cdots & \frac{2\left(
N-1\right) }{N+1} & \frac{N-1}{N+1} \\
\frac{1}{N+1} & \frac{2}{N+1} & \frac{3}{N+1} & \cdots & \frac{N-1}{N+1} &
\frac{N}{N+1}%
\end{array}%
\right)
\end{equation}%
capture many data on the $SU\left( N\right) $ symmetry; they give in
particular the expression of the simple roots $\mathbf{\alpha }_{1}\mathbf{%
,\alpha }_{2}\mathbf{,...,\alpha }_{N-1}$ in terms of the fundamental
weights $\mathbf{\omega }_{1},...,\mathbf{\omega }_{N-1}$ and vice versa;
that is $\mathbf{\alpha }_{i}=\sum_{j}K_{ij}\mathbf{\omega }_{j}$ and $%
\mathbf{\omega }_{i}=\sum_{j}K_{ij}^{-1}\mathbf{\alpha }_{j}$. Recall that
simple roots and fundamental weights obey the duality property $\mathbf{%
\alpha }_{i}.\mathbf{\omega }_{j}=\delta _{ij}$; we also have $\mathbf{%
\omega }_{i}.\mathbf{\omega }_{j}=K_{ij}^{-1}$.

\subsubsection{The lattice $\mathcal{L}_{su\left( N\right) }$}

\begin{figure}[tbph]
\centering
\hspace{0cm} \includegraphics[width=8cm]{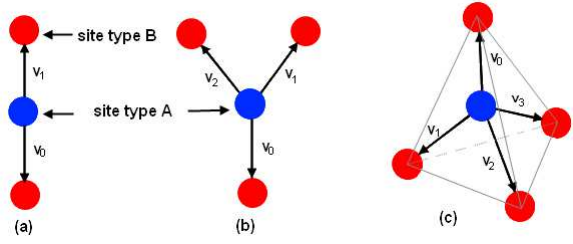}
\caption{ (a) 1A+2B lattice sites of $\mathcal{L}_{su\left(
2\right) }$;  A-type in blue and B-type in red; the 2B form a
$su\left( 2\right) $ \ doublet. (b) 1A+3B sites of $%
\mathcal{L}_{su\left( 3\right) }$; \ the 3B form a $su\left(
3\right) $\ triplet. (c) 1A+4B sites of $\mathcal{L}%
_{su\left( 4\right) }$  with 4B sites forming a regular
tetrahedron. }
\label{123}
\end{figure}

The lattice $\mathcal{L}_{su\left( N\right) }$ is a real $\left( N-1\right) $%
- dimensional crystal with two superposed integral sublattices $\mathcal{A}%
_{N}$ and $\mathcal{B}_{_{N}}$; each site $\mathbf{r}_{\mathbf{m}}$ of these
sublattices is generated by the $SU\left( N\right) $ simple roots $\mathbf{%
\alpha }_{1},...,\mathbf{\alpha }_{N-1}$;
\begin{equation}
\begin{tabular}{llll}
$\mathbf{r}_{\mathbf{m}}$ & $=$ & $m_{1}\mathbf{\alpha }_{1}+m_{2}\mathbf{%
\alpha }_{2}+...m_{N-1}\mathbf{\alpha }_{N-1}$ & ,%
\end{tabular}
\label{z1}
\end{equation}%
with $m_{i}$ integers; for illustration see the schema (a), (b), (c) of the
figure (\ref{123}) corresponding respectively to $N=2,3,4$; and which may be
put in one to one with the $sp^{1}$, $sp^{2}$ and $sp^{3}$ hybridization of
the carbon atom orbital $2s$ and $2p$.\newline
On each lattice site $\mathbf{r}_{m}$ of $\mathcal{L}_{su\left( N\right) }$;
say of A-type, lives a quantum state $A_{\mathbf{r}_{m}}$ coupled to the
nearest neighbor states; in particular the first nearest states $B_{\mathbf{r%
}_{m}+\mathbf{v}_{i}}$ and the second nearest ones $A_{\mathbf{r}_{m}+%
\mathbf{V}_{ij}}$. \newline
Generally, generic sites in $\mathcal{L}_{su\left( N\right) }$ have the
following properties:\newline
(\textbf{1}) $N$ first nearest neighbors with relative position vectors $%
\mathbf{v}_{i}$ constrained as
\begin{equation}
\begin{tabular}{lll}
$\mathbf{v}_{1}+\ldots +$ $\mathbf{v}_{N}$ & $=0$ & .%
\end{tabular}
\label{21}
\end{equation}%
These constraint relations are solved in terms of the $SU\left( N\right) $
weight vectors $\mathbf{\lambda }_{i}$ (resp. $-\mathbf{\lambda }_{i}$) of
the fundamental (anti-fundamental) representation as follows%
\begin{equation}
\begin{tabular}{lllll}
$\mathbf{v}_{i}$ & $=a\mathbf{\lambda }_{i}$ & $\equiv $ & $d\frac{\mathbf{%
\lambda }_{i}}{\left\Vert \mathbf{\lambda }_{i}\right\Vert }$ & ,%
\end{tabular}
\label{22}
\end{equation}%
where $d$ is the relative distance between the closest $\mathcal{L}%
_{su\left( N\right) }$ sites. The $\mathbf{\lambda }_{i}$'s which satisfy $%
\mathbf{\lambda }_{1}+\ldots +$ $\mathbf{\lambda }_{N}=0$ can be nicely
expressed in terms of the fundamental weights $\mathbf{\omega }_{1},...,%
\mathbf{\omega }_{N-1}$ as follows%
\begin{equation}
\begin{tabular}{llll}
$\mathbf{\lambda }_{1}=\mathbf{\omega }_{1},$ & $\mathbf{\lambda }_{i}=%
\mathbf{\omega }_{i}-\mathbf{\omega }_{i-1},$ & $\mathbf{\lambda }_{N}=-%
\mathbf{\omega }_{N-1}$ & .%
\end{tabular}
\label{ld}
\end{equation}%
From the QFT view, this means that the quantum states at $\mathbf{r}_{m}+%
\mathbf{v}_{i}$ sites are labeled by the $\mathbf{\lambda }_{i}$ weights as $%
B_{\mathbf{r}_{m}+\mathbf{v}_{i}}\equiv B_{\mathbf{\lambda }_{i}}\left(
\mathbf{r}_{m}\right) $ and so the multiplet%
\begin{equation}
\begin{tabular}{ll}
$\left(
\begin{array}{c}
|\mathbf{\lambda }_{1}> \\
\vdots \\
|\mathbf{\lambda }_{N}>%
\end{array}%
\right) \equiv $ \ \underline{$\mathbf{N}$}, & $\qquad \mathbf{\lambda }%
_{1}+\ldots +\mathbf{\lambda }_{N}=0,$%
\end{tabular}%
\end{equation}%
transform in the fundamental representation of $SU\left( N\right) $. \newline
(\textbf{2}) $N\left( N-1\right) $ second nearest neighbors of A-type with
relative position vectors $\mathbf{V}_{ij}$ given by $\mathbf{v}_{i}-\mathbf{%
v}_{j}$ and obeying the constraint relation $\sum_{i,j}\mathbf{V}_{ij}=0.$
This condition is naturally solved by (\ref{21}) and (\ref{22}) showing that
the relative vectors between second nearest neighbors are proportional to $%
SU\left( N\right) $ roots $\mathbf{\beta }_{ij}$ like
\begin{equation}
\begin{tabular}{lll}
$\mathbf{V}_{ij}=a\mathbf{\beta }_{ij}$ & ,\qquad & $\mathbf{\beta }_{ij}=%
\mathbf{\lambda }_{i}-\mathbf{\lambda }_{j},$%
\end{tabular}
\label{26}
\end{equation}%
and so the condition $\sum \mathbf{V}_{ij}=0$ turns to a $SU\left( N\right) $
property on its adjoint representation labeled by the roots.

\subsection{Energy dispersion relation}

Restricting the analysis to the first nearest neighbors described by eq(\ref%
{HN}) in the limit $t^{\prime }\rightarrow 0$, the hamiltonian $H_{N}$ on $%
\mathcal{L}_{su\left( N\right) }$ reduces to%
\begin{equation}
\begin{tabular}{lll}
$H_{N}=$ & $-t\sum\limits_{\mathbf{r}_{i}}\left( \sum\limits_{n=1}^{N}a_{%
\mathbf{r}_{i}}b_{\mathbf{r}_{i}{\scriptsize +}\mathbf{v}_{n}}^{\dagger
}\right) +hc$ & ,%
\end{tabular}
\label{B1}
\end{equation}%
where now $\mathbf{r}_{i}$ and $\mathbf{v}_{n}$ are $\left( N-1\right) $-
dimensional vectors. By using the Fourier transform of the field operators $%
a_{\mathbf{r}_{i}}$ and $B_{\mathbf{r}_{m}+\mathbf{v}_{i}}^{\pm }$ namely,%
\begin{equation}
\begin{tabular}{llll}
$a_{\mathbf{r}_{i}}\sim \dsum\limits_{\mathbf{k}}e^{i\mathbf{k.r}_{m}}a_{%
\mathbf{k}}^{\pm }$ & , & $b_{\mathbf{r}_{m}+\mathbf{v}_{i}}\sim
\dsum\limits_{\mathbf{k}}e^{i\mathbf{k.}\left( \mathbf{r}_{m}+\mathbf{v}%
_{i}\right) }b_{\mathbf{k}}$ &
\end{tabular}%
\end{equation}%
we can put the hamiltonian $H_{N}$ as a sum over the wave vectors $\mathbf{k}
$ in the following way;
\begin{equation}
\begin{tabular}{ll}
$H_{N}=\dsum\limits_{\mathbf{k}}\left( a_{\mathbf{k}}^{\dagger },b_{\mathbf{k%
}}^{\dagger }\right) \left(
\begin{array}{cc}
0 & \varepsilon _{\mathbf{k}} \\
\varepsilon _{\mathbf{k}}^{\ast } & 0%
\end{array}%
\right) \left(
\begin{array}{c}
a_{\mathbf{k}} \\
b_{\mathbf{k}}%
\end{array}%
\right) $ & ,%
\end{tabular}%
\end{equation}%
with $\varepsilon _{\mathbf{k}}=t\sum_{i}e^{ia\mathbf{k}.\mathbf{\lambda }%
_{i}}$. This complex number can be also written as $t\sum_{i}e^{ia\mathbf{k}%
.\left( \mathbf{\omega }_{i}-\omega _{i-1}\right) }$ with $\omega
_{-1}=0=\omega _{N}$. The energy dispersion relation of the "valence" and
"conducting" bands are obtained by diagonalizing the hamiltonian $H_{N}$;
they are given by $\pm \left\vert \varepsilon _{\mathbf{k}}\right\vert $
with,%
\begin{equation}
\begin{tabular}{ll}
$\left\vert \varepsilon _{\mathbf{k}}\right\vert =t\sqrt{N+2\dsum%
\limits_{i<j=1}^{N}\cos \left[ a\mathbf{k}.\left( \mathbf{\lambda }_{i}%
\mathbf{-\lambda }_{j}\right) \right] }$ & .%
\end{tabular}
\label{di}
\end{equation}%
Notice that $\left\vert \varepsilon _{\mathbf{k}}\right\vert $ depends
remarkably in the difference of the weights $\mathbf{\lambda }_{i}\mathbf{%
-\lambda }_{j}$; which by help of eq(\ref{ld}), can be completely expressed
in terms of the fundamental weights. \newline
To get the Fermi wave vectors $\mathbf{k}_{F}$ for which the oscillating
multi-variable function $\varepsilon _{\mathbf{k}}=t\sum_{l}e^{ia\mathbf{k}.%
\mathbf{\lambda }_{l}}$ vanish, we will proceed as follows: First, we work
out an explicit example; then we give the general result. To that purpose,
we expand the wave vector $\mathbf{k}$ in the weight vector basis as
follows,
\begin{equation}
\begin{tabular}{lll}
$\mathbf{k}=\dsum\limits_{i=1}^{N-1}Q_{i}\mathbf{\omega }_{i},$ & $\left(
Q_{1},...,Q_{N}\right) \in \mathbb{R}^{N}$ & ,%
\end{tabular}
\label{kw}
\end{equation}%
and focus on working out the solution for the particular case where all the $%
Q_{i}$'s are equal, i.e: $Q_{1}=Q_{2}=...=Q_{N-1}=Q.$ General solutions are
obtained from this particular case by performing lattice translations along
the $\mathbf{\omega }_{i}$-directions; this leads to the new values $Q_{l}=Q+%
\frac{2\pi n_{l}}{N}$ with $n_{l}$ integers. Obviously, one may also expand
the wave vector $\mathbf{k}$ like
\begin{equation}
\begin{tabular}{llll}
$\mathbf{k=}\dsum\limits_{i=1}^{N-1}k_{i}\mathbf{\alpha }_{i}$ & , & $\left(
k_{1},...,k_{N}\right) \in \mathbb{R}^{N}$ & .%
\end{tabular}
\label{kal}
\end{equation}%
But this is equivalent to (\ref{kw}); the relation between the $Q_{l}$'s and
the $k_{l}$'s is obtained by substituting $\mathbf{\alpha }%
_{i}=\sum_{j}K_{ij}\mathbf{\omega }_{j}$ into (\ref{kal}) and identifying it
with (\ref{kw}). To compute the factors $e^{ia\mathbf{k}.\mathbf{\lambda }%
_{l}}$, we express the vectors $\mathbf{\lambda }_{l}$ in terms of the
simple roots as follows%
\begin{equation}
\begin{tabular}{llll}
$\mathbf{\lambda }_{1}=\mathbf{\omega }_{1},$ & $\mathbf{\lambda }_{2}=%
\mathbf{\omega }_{1}-\mathbf{\alpha }_{1},$ & $\mathbf{\ldots }$ & $\mathbf{%
\lambda }_{N}=\mathbf{\omega }_{1}-\mathbf{\alpha }_{1}-...-\mathbf{\alpha }%
_{_{N-1}},$%
\end{tabular}%
\end{equation}%
then we use the root/weight duality relation $\mathbf{\omega }_{i}.\mathbf{%
\alpha }_{j}=\delta _{ij}$ as well as the simple choice $Q_{l}=Q$ to put the
scalar product $\mathbf{k}.\mathbf{\lambda }_{l}$ into the following form $%
\left( \mathbf{k}.\mathbf{\lambda }_{l}\right) =\left( \mathbf{k}.\mathbf{%
\omega }_{1}\right) -lQ$, $l=1,...,N-1.$ \ Putting this expression back into
$\varepsilon _{\mathbf{k}}$ and setting $\xi =e^{iaQ}$, we obtain $%
\varepsilon _{\mathbf{k}}=e^{ia\mathbf{k}.\mathbf{\omega }_{1}}\left[ 1+\xi
+...+\xi ^{N-1}\right] =0$ which is exactly solved by the \emph{N-th} roots
of unity namely%
\begin{equation}
\begin{tabular}{lllll}
$Q=\pm \frac{2s\pi }{aN}$ & , & $s=1,...,\left[ \frac{N}{2}\right] $ & . &
\end{tabular}%
\end{equation}%
Therefore the Dirac points are, up to lattice translations, located at the
wave vectors $\mathbf{k}_{s}=\pm \frac{2s\pi }{aN}\sum_{i}\mathbf{\omega }%
_{i}=\pm \frac{2s\pi }{aN}\sum_{i}\mathbf{\alpha }_{i}.$

\section{Leading models}

In this section, we study the cases $N=2,4$ as $N=3$ corresponds precisely
to the \emph{2D} graphene before. The case $N=5$ will be studied in the next
section seen its remarkable relation with \emph{4D}\ lattice QCD.

\subsection{The $su\left( 2\right) $ model}

\begin{figure}[tbph]
\centering
\hspace{0cm} \includegraphics[width=8cm]{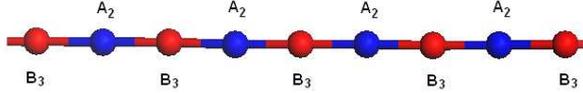}
\caption{{ the lattice }$\mathcal{L}_{su\left( 2\right) }$%
{ \ given by the superposition of two sublattices }$\mathcal{A}%
_{su\left( 2\right) }${ \ (in blue) and }$\mathcal{B}%
_{su\left( 2\right) }$ { (in red). The atoms may be thought of
as carbons in the }${ sp}^{{ 1}}${ %
\ hybridization state.}}
\label{ch}
\end{figure}

In this case, the lattice $\mathcal{L}_{su\left( 2\right) }$, which is
depicted in the figure (\ref{ch}), is a one dimensional chain with
coordinate positions $\mathbf{r}_{m}=ma$ where $a$ is the site spacing and $%
m $ an arbitrary integer.

Examples of carbon chains with delocalized electrons are given by one of the
three following molecules%
\begin{equation}
\begin{tabular}{l|l|l}
\ \ \ \ {\small chain} & \ \ \ \ \ \ \ \ \ \ \ \ {\small molecule} & {\small %
delocalized electrons} \\ \hline
{\small polyacetylene} & ${\small ...-CH=CH-CH=CH-CH-...}$ & {\small \ \ \ \
\ \ \ \ \ 1} \\
{\small cumulene} & ${\small ...=C=C=C=C=C=...}$ & {\small \ \ \ \ \ \ \ \ \
2} \\
{\small poly-yne} & ${\small ...-C}\equiv {\small C-C}\equiv {\small C-C}%
\equiv {\small C-...}$ & {\small \ \ \ \ \ \ \ \ \ 2} \\ \hline
\end{tabular}%
\end{equation}%
These molecules can be taken as {the graphene }bridge ultimately narrowed
down to a few- carbon atoms or a single-atom width \textrm{\citep{H1,H2,H4}}%
. Each site of $\mathcal{L}_{su\left( 2\right) }$ has two first nearest
neighbors forming an $su\left( 2\right) $ doublet; and two second nearest
ones that are associated with the two roots $\pm \alpha $ of $su\left(
2\right) $ in agreement with the generic result summarized in the table,%
\begin{equation}
\begin{tabular}{l|l|l|l|ll}
{\small nearest neighbors} & $SU\left( N\right) $ & $SU\left( 2\right) $ & $%
SU\left( 3\right) $ & $SU\left( 4\right) $ & $SU\left( 5\right) $ \\ \hline
{\small \ \ \ \ \ \ \ \ \ \ \ first } & $N$ & $2$ & $3$ & $4$ & $5$ \\
{\small \ \ \ \ \ \ \ \ \ \ \ second} & $N\left( N-1\right) $ & $2$ & $6$ & $%
12$ & $20$ \\ \hline
\end{tabular}%
\end{equation}%
In the $SU\left( 2\right) $ lattice model, eqs(\ref{v}) read as
\begin{equation}
\begin{tabular}{lll}
$\mathbf{v}_{0}+\mathbf{v}_{1}=0$ & , & (a) \\
$\mathbf{V}_{01}=\mathbf{v}_{0}-\mathbf{v}_{1}$ & , & (b)%
\end{tabular}%
\end{equation}%
and are solved by the fundamental weights $\lambda _{1}=+\frac{1}{2},$ $%
\lambda _{2}=-\frac{1}{2}$ of the $SU\left( 2\right) $ fundamental
representation; i.e the isodoublet.

\begin{figure}[tbph]
\centering
\hspace{0cm} \includegraphics[width=6cm]{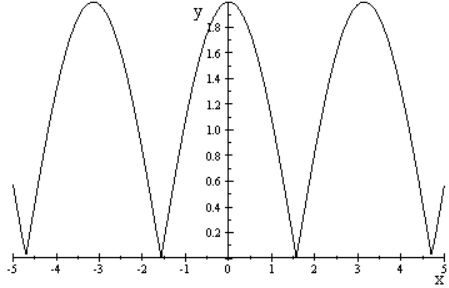}
\caption{{ Energy dispersion relation of 1D poly-acetylene
chain.}}
\label{2N}
\end{figure}

\emph{1) polyacetylene}\newline
The hamiltonian of the polyacetylene, where each carbon has one delocalized
electron, is given by
\begin{equation}
\begin{tabular}{lll}
$H_{t}$ & $=-t\dsum\limits_{m}\left( a_{\mathbf{r}_{m}}b_{\mathbf{r}%
_{m}+a}^{+}+a_{\mathbf{r}_{m}}b_{\mathbf{r}_{m}-a}^{+}\right) +hc$ & .%
\end{tabular}%
\end{equation}%
Substituting $N=2$ in (\ref{di}), we get the following energy dispersion
relation%
\begin{equation}
\left\vert \varepsilon _{k}\right\vert =t\sqrt{2+2\cos \left( 2ak\right) }
\end{equation}%
which is also equal to $2t\cos \left( ka\right) $ in agreement with the
expression $\varepsilon _{k}=t\left( e^{iak}+e^{-iak}\right) $; see also
figure (\ref{2N}). Moreover, the vanishing condition $\varepsilon _{2}\left(
k\right) =0$ is solved by the wave vectors $k_{\pm }=\pm \frac{\pi }{2a}$ $%
\func{mod}\left( \frac{2\pi }{a}\right) $.

\emph{2) cumulene and poly-yne}\newline
In the case of cumulene and poly-yne, the two delocalized electrons are
described by two wave functions $\phi _{_{\mathbf{r}_{m}}}^{1}$, $\phi _{_{%
\mathbf{r}_{m}}}^{2}$. The tight binding hamiltonian modeling the hopping of
these electrons is a generalization of $H_{t}$. Let $a_{\mathbf{r}%
_{m}}^{\alpha }$, $a_{\mathbf{r}_{m}}^{\alpha +}$, $\alpha =1,2$ (resp. $b_{%
\mathbf{r}_{m}\pm a}^{\alpha }$, $b_{\mathbf{r}_{m}\pm a}^{\alpha +}$) be
the annihilation and creation operators at the site $\mathbf{r}_{m}$ (resp. $%
\mathbf{r}_{m}\pm a$), the hamiltonian reads as follows%
\begin{equation}
\begin{tabular}{lll}
$H_{t,t^{\prime }}$ & $=-\dsum\limits_{m}\dsum\limits_{\alpha ,\beta
=1}^{2}\left( a_{\mathbf{r}_{m}}^{\alpha }t_{\alpha \beta }b_{\mathbf{r}%
_{m}+a}^{\beta +}+a_{\mathbf{r}_{m}}^{\alpha }t_{\alpha \beta }^{\prime }b_{%
\mathbf{r}_{m}-a}^{\beta +}\right) +hc$ & ,%
\end{tabular}%
\end{equation}%
where $t_{\alpha \beta }$ and $t_{\alpha \beta }^{\prime }$ are hop energy
matrices which are identical for cumulene ($t_{\alpha \beta }=t_{\alpha
\beta }^{\prime }$), but different for poly-yne ($t_{\alpha \beta }\neq
t_{\alpha \beta }^{\prime }$). Mapping this hamiltonian to the reciprocal
space, we get%
\begin{equation}
\begin{tabular}{lll}
$H_{t,t^{\prime }}$ & $=-2\dsum\limits_{m}\dsum\limits_{\alpha ,\beta
=1}^{2}\left( a_{k}^{1},b_{k}^{1},a_{k}^{2},b_{k}^{2}\right) \left(
\begin{array}{cccc}
0 & A & 0 & B \\
A^{\ast } & 0 & C^{\ast } & 0 \\
0 & C & 0 & D \\
B^{\ast } & 0 & D^{\ast } & 0%
\end{array}%
\right) \left(
\begin{array}{c}
a_{k}^{1+} \\
b_{k}^{1+} \\
a_{k}^{2+} \\
b_{k}^{2+}%
\end{array}%
\right) $ & ,%
\end{tabular}%
\end{equation}%
with%
\begin{equation}
\begin{tabular}{lll}
$A\left( k\right) =$ & $t_{11}e^{ika}+t_{11}^{\prime }e^{-ika}$ &  \\
$B\left( k\right) =$ & $t_{12}e^{ika}+t_{12}^{\prime }e^{-ika}$ &  \\
$C\left( k\right) =$ & $t_{21}e^{ika}+t_{21}^{\prime }e^{-ika}$ &  \\
$D\left( k\right) =$ & $t_{22}e^{ika}+t_{22}^{\prime }e^{-ika}$ &
\end{tabular}%
\end{equation}%
Now, using the fact that the two delocalized electrons are
indistinguishable, it is natural to assume the following relations on the
hop energies $t_{11}=t_{22}$, $t_{12}=t_{21}$ and the same thing for the $%
t_{\alpha \beta }^{\prime }$ matrix. This leads to the relations $A=D$, $B=C$
and so the above hamiltonian simplifies. In this case, the four energy
eigenvalues are given by
\begin{equation}
\begin{tabular}{lll}
$E_{\pm }=$ & $\pm \sqrt{\left( A^{\ast }+B^{\ast }\right) \left( A+B\right)
}$ & , \\
$E_{\pm }^{\prime }=$ & $\pm \sqrt{\left( A^{\ast }-B^{\ast }\right) \left(
A-B\right) }$ & ,%
\end{tabular}%
\end{equation}%
and the zeros modes are given by $e^{2ika}=-\frac{t_{11}^{\prime }}{t_{11}}=-%
\frac{t_{12}^{\prime }}{t_{12}}$. Since in the case of cumulene we have $%
t_{\alpha \beta }=t_{\alpha \beta }^{\prime }$, \ it follows that the zero
modes are located as $k=\pm \frac{\pi }{2a}$ $\func{mod}\frac{2\pi }{a}$.

\begin{figure}[tbph]
\centering
\hspace{0cm} \includegraphics[width=8cm]{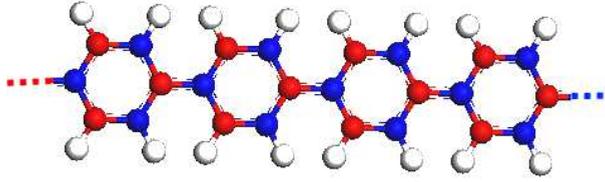}
\caption{{ a periodic chain in 3D space with unit cells given
by hexagonal cycles. Each cycle has six delocalized electrons. }}
\label{cha}
\end{figure}

\emph{3) nanoruban}\newline
We end this paragraph noting that such analysis may be also extended to the
particular case of the periodic chain made by the junction of hexagonal
cycles as depicted in the figure (\ref{cha}).
This chain, which can be also interpreted as the smallest graphene
nanoruban, is very particular from several issues; first its unit cells can
be taken as given by the hexagonal cycles; second amongst the \emph{6}
carbons of the unit cycle, \emph{4} of them have two first nearest neighbors
and the \emph{2} others have three first nearest ones. The third
particularity is that the tight binding description of this chain is somehow
more complicated with respect to the previous examples. Below we focus on
the electronic properties of a given cycle by using the same approach we
have been considering in this study.

\subsection{Kekulé cycles}

\begin{figure}[tbph]
\begin{center}
\hspace{0cm} \includegraphics[width=10cm]{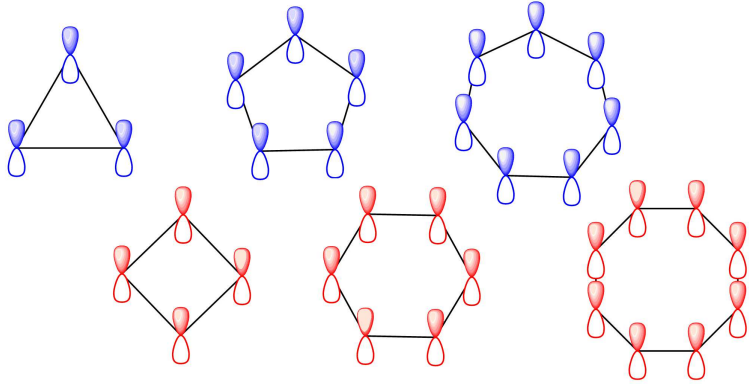}
\end{center}
\par
\caption{{ Six examples of Kekulé cycles type C}$_{%
{ n}}${ H}$_{{ n}}${ %
\ with }$n=3,4,5,6,7,8${ . The cations C}$^{+}${ %
\ of these molecules form a heavy skeleton represented by n-polygons. The
orbitals in the normal direction are associated with the delocalized
Pi-electrons.}}
\label{F1}
\end{figure}

\begin{figure}[tbph]
\centering
\hspace{0cm} \includegraphics[width=8cm]{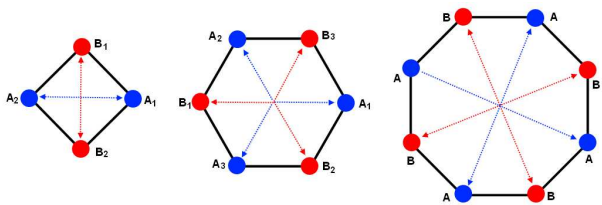}
\caption{{ Kekulé molecules as the superposition of two
sublattices. Sublattice }$A${ \ in blue and sublattice }$%
\mathcal{B}$ { in red. Except the benzene, these molecules are
generally are non planar.}}
\label{F}
\end{figure}

Kekulé cycles are organic molecules named in honor to the German chemist
Friedrich Kekulé known for his works on tetravalent structure of carbon and
the cyclic structure of benzene $C_{6}H_{6}$. These molecules; in particular
the family $C_{n}H_{n}$ with $n\geq 3$, may be thought of as one dimensional
cycles living in the \emph{3D} space; they involve carbon atoms (eventually
other atoms such as Nitrogen) arranged in a cyclic lattice with both $\sigma
$- and $\pi $-bonds. All these carbon atoms are in the \emph{sp}$^{2}$
hybridization; they have \emph{3n}$\ $covalent $\sigma $-bonds defining a
\emph{quasi-planar} skeleton; and \emph{n} delocalized $\pi $-bonds with Pi
electron orbital expanding in the normal direction as shown in the examples
of \textrm{fig(\ref{F1})}.
Our interest into Kekulé molecules, in particular to the $C_{2N}H_{2N}$
family, comes from the fact that they can be viewed as the \emph{1D}\
analogue of the\emph{\ 2D} graphene monolayer; they may be also obtained
from the poly-acetylene chain by gluing the ends. It is then interesting to
explore the electronic properties of this special class of systems by using
the tight binding model and symmetries. To illustrate the method, we focus
on the benzene $C_{6}H_{6}$ thought of as the superposition of two $%
C_{3}H_{3}$ sub-molecules as depicted in figure (\ref{F}).
From group theory view, the positions $\mathbf{v}_{1},$ $\mathbf{v}_{2},$ $%
\mathbf{v}_{3},$ $\mathbf{v}_{4},$ $\mathbf{v}_{5},$ $\mathbf{v}_{6}$ of the
carbon atoms are given by the six roots of the $SU\left( 3\right) $
symmetry.; that is
\begin{equation}
\begin{tabular}{llll}
$\mathbf{v}_{i}=a\mathbf{\alpha }_{i}$ & , & $\mathbf{v}_{3+i}=-a\mathbf{%
\alpha }_{i}$ & ,%
\end{tabular}%
\end{equation}%
where $a=1.39$ $\mathring{A}$ and where the three $\mathbf{\alpha }_{i}$'s
are as in section 2.

\begin{figure}[tbph]
\centering
 \includegraphics[width=6cm]{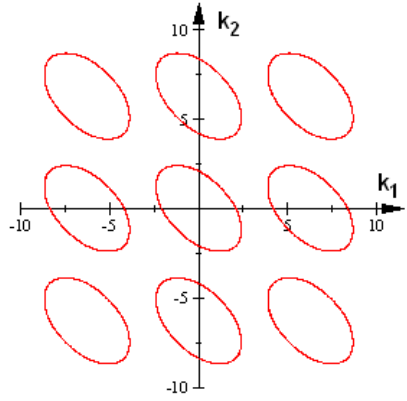}
\caption{{ the plot of the energy dispersion relation }$%
\protect\varepsilon _{k_{1},k_{2}}${ \ and the constraint
relation }$\protect\eta _{k_{1},k_{2}}${ \ }$=${ %
\ }$\sin k_{1}${ \ }$+\sin k_{2}${ \ }$-\sin
\left( k_{1}+k_{2}\right) ${ \ }$=0${ \ in the
reciprocal space.}}
\label{6F}
\end{figure}

\emph{tight binding description}\newline
The electronic properties of the $C_{6}H_{6}$ are captured by the
pi-electrons of the carbons. Denoting by $\mathbf{a}_{\mathbf{\mathbf{r}_{i}}%
}^{\dagger }$, $\mathbf{a}_{\mathbf{\mathbf{r}_{i}}}$ (resp. $\mathbf{b}_{%
\mathbf{\mathbf{r}_{i}}}^{\dagger }$, $\mathbf{b}_{\mathbf{\mathbf{r}_{i}}}$%
) the usual electronic creation and annihilation operators associated with
the A$_{i}$ (B$_{j}$) atoms in the sublattice $\mathcal{A}_{benz}$ ($%
\mathcal{B}_{benz}$), the tight binding hamiltonian of the benzene,
restricted to first nearest neighbors, reads as follows,
\begin{equation}
\begin{tabular}{lll}
$H_{benz}=$ & $-t\sum\limits_{m=-\infty }^{\infty
}\sum\limits_{l=1}^{3}\left( \sum\limits_{j=1}^{2}\mathbf{a}_{\mathbf{r}%
_{l,m}}\mathbf{b}_{\mathbf{r}_{l,m}+\mathbf{v}_{l,j}}^{\dagger }\right) +hc$
& .%
\end{tabular}
\label{tb}
\end{equation}%
In this relation, the position vectors $\mathbf{r}_{lm}$ have two indices; $%
l $ and $m$. The first one takes the values $l=1,2,3$; it indexes the three
atoms in $\mathcal{A}_{benz}$; and the three ones in $\mathcal{B}_{benz}$.
These positions are as follows,
\begin{equation}
\begin{tabular}{llll}
$\mathbf{r}_{2l-1,m}^{A}=\mathbf{r}_{1m}^{A},$ $\mathbf{r}_{3m}^{A},$ $%
\mathbf{r}_{5m}^{A}$ & , & $\mathbf{r}_{2l,m}^{B}=\mathbf{r}_{2m}^{B},$ $%
\mathbf{r}_{4m}^{B},$ $\mathbf{r}_{6m}^{B}$ & .%
\end{tabular}%
\end{equation}%
The second integer is an arbitrary number ($m\in \mathbb{Z}$); it captures
the periodicity of the cycle and encodes in some sense the rotational
invariance with respect to the axis of the planar molecule. \newline
To fix the ideas, think about $\mathbf{r}_{lm}$ as the \emph{l-th} electron
in the sublattice $\mathcal{A}_{benz}$; that is $\mathbf{r}_{lm}\equiv
\mathbf{r}_{2l-1,m}^{A}$. After a hop of this electron to the two first
nearest carbons in $\mathcal{B}_{benz}$, the new position is
\begin{equation}
\begin{tabular}{llll}
$\mathbf{r}_{2l,m}^{B}=\mathbf{r}_{lm}+\mathbf{v}_{lj}$ & , & $j=\pm $ & .%
\end{tabular}
\label{del}
\end{equation}%
\ where the $\mathbf{v}_{lj}$s are the relative positions of the first
nearest neighbors. \newline
Taking the Fourier transform of the creation and annihilation operators, $c_{%
\mathbf{r}_{n}}^{\pm }=\sum_{k}e^{\pm i\mathbf{k.r}_{n}}c_{\mathbf{k}}^{\pm
} $ with $c_{\mathbf{k}}^{\pm }$\ standing for $a_{\mathbf{k}}^{\pm }$, $b_{%
\mathbf{k}}^{\pm }$, we get an expression involving the product of three
sums $\sum_{\mathbf{k}}$ $\sum_{\mathbf{k}^{\prime }}$ $\sum_{m}$. Then,
using the discrete rotational invariance with respect to the axis of the
molecule, we can eliminate the sum\textrm{\ }$\sum_{m}$ in terms of a Dirac
delta function $\delta _{2}\left( \mathbf{k}-\mathbf{k}^{\prime }\right) $
and end, after integration with respect $\mathbf{k}^{\prime }$, with the
following result,
\begin{equation}
\begin{tabular}{ll}
$\mathcal{H}_{benz}=\sum\limits_{\mathbf{k}}\text{ }\left( a_{\mathbf{k}},b_{%
\mathbf{k}}\right) \left(
\begin{array}{cc}
0 & \varepsilon _{\mathbf{k}} \\
\varepsilon _{\mathbf{k}}^{\ast } & 0%
\end{array}%
\right) \left(
\begin{array}{c}
a_{\mathbf{k}}^{\dagger } \\
b_{\mathbf{k}}^{\dagger }%
\end{array}%
\right) $ & .%
\end{tabular}
\label{mat}
\end{equation}%
with
\begin{equation}
\begin{tabular}{ll}
$\varepsilon _{\mathbf{k}}=-t\sum\limits_{l=1}^{3}\left( e^{i\mathbf{k}.%
\mathbf{v}_{l-}}+e^{i\mathbf{k}.\mathbf{v}_{l+}}\right) $ & .%
\end{tabular}
\label{an}
\end{equation}%
Notice that like in graphene, the above hamiltonian has two eigenvalues $\pm
\left\vert \varepsilon _{\mathbf{k}}\right\vert $. Moreover, substituting
the $\mathbf{v}_{l\pm }$'s by their explicit expressions in terms of the $%
SU\left( 3\right) $ roots $\mathbf{\alpha }_{l}$, we obtain the following
dispersion relation together with a constraint relation capturing the
planarity property of the molecule%
\begin{equation}
\begin{tabular}{lll}
$\varepsilon _{\mathbf{k}}=-2t\sum\limits_{l=0}^{2}\cos \left( \frac{a\sqrt{2%
}}{2}\mathbf{k}.\mathbf{\alpha }_{l}\right) ,$ & $\eta _{\mathbf{k}%
}=\sum\limits_{l=0}^{2}\sin \left( \frac{a\sqrt{2}}{2}\mathbf{k}.\mathbf{%
\alpha }_{l}\right) =0$ & .%
\end{tabular}
\label{en}
\end{equation}%
Notice that the constraint equation $\eta _{\mathbf{k}}=0$ allows us to
express the $k_{2}$ component of the wave vector in terms of $k_{1}$ and
vice versa as depicted in fig(\ref{6F}). This relation plays a crucial role
in the determination of the wave vectors at the Fermi level.

\subsection{The diamond model}

The diamond model lives on the lattice $\mathcal{L}_{su\left( 4\right) }$;
this is a 3-dimensional crystal given by the superposition of two isomorphic
sublattices $\mathcal{A}_{4}$ and $\mathcal{B}_{4}$ along the same logic as
in the case of the \emph{2D} honeycomb. Each site $\mathbf{r}_{m}$ in $%
\mathcal{L}_{su\left( 4\right) }$ has \emph{4} first nearest neighbors at $%
\left( \mathbf{r}_{m}+\mathbf{v}_{i}\right) $ forming the vertices of a
regular tetrahedron. A way to parameterize the relative positions $\mathbf{v}%
_{i}$ with respect to the central position at $\mathbf{r}_{m}$ is to embed
the tetrahedron inside a cube; in this case we have:%
\begin{equation}
\begin{tabular}{llll}
$\mathbf{v}_{1}=\frac{d}{\sqrt{3}}\left( -1,-1,+1\right) $ & , & $\mathbf{v}%
_{2}=\frac{d}{\sqrt{3}}\left( -1,+1,-1\right) $ &  \\
$\mathbf{v}_{3}=\frac{d}{\sqrt{3}}\left( +1,-1,-1\right) $ & , & $\mathbf{v}%
_{0}=\frac{d}{\sqrt{3}}\left( +1,+1,+1\right) $ &
\end{tabular}
\label{vi}
\end{equation}%
Clearly these vectors satisfy the constraint relation $\mathbf{v}_{0}+%
\mathbf{v}_{1}+\mathbf{v}_{2}+\mathbf{v}_{3}=0$. Having these expressions,
we can also build the explicit positions of the \emph{12} second nearest
neighbors; these are given by $\mathbf{V}_{ij}=\mathbf{v}_{i}-\mathbf{v}_{j}$%
; but are completely generated by the following basis vectors
\begin{equation}
\begin{tabular}{lllll}
$\mathbf{R}_{1}=\frac{d}{\sqrt{3}}\left( 2,2,0\right) $ & , & $\mathbf{R}%
_{2}=\frac{d}{\sqrt{3}}\left( 0,-2,2\right) $ & , & $\mathbf{R}_{3}=\frac{d}{%
\sqrt{3}}\left( -2,2,0\right) $%
\end{tabular}
\label{RR}
\end{equation}%
that are related to $\mathbf{V}_{ij}$ as $\mathbf{R}_{i}=\mathbf{V}_{\left(
i-1\right) i}$. We also have:

\begin{itemize}
\item the intersection matrix of the $\mathbf{R}_{i}$ vectors
\begin{equation}
\mathbf{R}_{i}.\mathbf{R}_{j}=\frac{4d^{2}}{3}\mathbf{K}_{ij}  \label{kij}
\end{equation}%
with
\begin{equation}
\begin{tabular}{llll}
$\mathbf{K}_{ij}=\left(
\begin{array}{ccc}
2 & -1 & 0 \\
-1 & 2 & -1 \\
0 & -1 & 2%
\end{array}%
\right) $ & , & $\mathbf{K}_{ij}^{-1}=\left(
\begin{array}{ccc}
\frac{3}{4} & \frac{2}{4} & \frac{1}{4} \\
\frac{2}{4} & \frac{4}{4} & \frac{2}{4} \\
\frac{1}{4} & \frac{2}{4} & \frac{3}{4}%
\end{array}%
\right) $ &
\end{tabular}%
\end{equation}

\item the special relation linking the $\mathbf{R}_{i}$'s and $\mathbf{v}%
_{0} $,%
\begin{equation}
\begin{tabular}{ll}
$\frac{3}{4}\mathbf{R}_{1}+\frac{2}{4}\mathbf{R}_{2}+\frac{1}{4}\mathbf{R}%
_{3}=\mathbf{v}_{0}$ & .%
\end{tabular}
\label{v0}
\end{equation}
\end{itemize}

Concerning the vector positions of the remaining \emph{9} second neighbors,
\emph{3} of them are given by $-R_{1},-R_{2},-R_{3}$ and the other \emph{6}
by the linear combinations $R_{4}=V_{02},$ $R_{5}=V_{13}$, $R_{6}=V_{03}$
with
\begin{equation}
\begin{tabular}{lll}
$V_{02}=R_{1}+R_{2},$ & $V_{13}=R_{2}+R_{3},$ & $V_{03}=R_{1}+R_{2}+R_{3}$.%
\end{tabular}
\label{R}
\end{equation}%
From this construction, it follows that generic positions $\mathbf{r}_{%
\mathbf{m}}^{A}\equiv \mathbf{r}_{\mathbf{m}}$ and $\mathbf{r}_{\mathbf{m}%
}^{B}$ in the $\mathcal{A}_{4}$ and $\mathcal{B}_{4}$ sublattices are given
by%
\begin{equation}
\begin{tabular}{llllll}
$\mathcal{A}_{_{4}}$ & : & $\mathbf{r}_{\mathbf{m}}$ & $=$ & $m_{1}\mathbf{R}%
_{1}+m_{2}\mathbf{R}_{2}+m_{3}\mathbf{R}_{3}$ & , \\
$\mathcal{B}_{_{4}}$ & : & $\mathbf{r}_{\mathbf{m}}^{B}$ & $=$ & $\mathbf{r}%
_{\mathbf{m}}+\mathbf{v}$ & ,%
\end{tabular}
\label{mr}
\end{equation}%
where $\mathbf{m}=\left( m_{1}\mathbf{,}m_{2}\mathbf{,}m_{3}\right) $ is an
integer vector and where the shift vector $\mathbf{v=r}_{\mathbf{m}}^{B}-%
\mathbf{r}_{\mathbf{m}}^{A}$ is one of $\mathbf{v}_{i}$'s in (\ref{vi}).

\emph{1)} \emph{Energy dispersion relation}\newline
First notice that as far as the electronic properties are concerned, the
figures (a), (b), (c) of (\ref{123}) are respectively associated with the $%
sp^{1}$, $sp^{2}$ and $sp^{3}$ hybridizations of the atom orbitals; i.e:%
\begin{equation}
\begin{tabular}{l|l|l}
{\small \ \ \ figures} & {\small hybridization} & {\small example of
molecules} \\ \hline
\ \ {\small (\ref{123}-a)} & ${\small \ \ \ \ sp}^{1}$ & {\small acetylene}
\\
\ \ {\small (\ref{123}-b)} & ${\small \ \ \ \ sp}^{2}$ & {\small graphene}
\\
\ \ {\small (\ref{123}-c)} & ${\small \ \ \ \ sp}^{3}$ & {\small diamond} \\
\hline
\end{tabular}%
\end{equation}%
In (\ref{123}-a) and\ (\ref{123}-b),{\small \ }the atoms have delocalized
pi- electrons that capture the electronic properties of the lattice atoms
and have the following dispersion relation,
\vspace{2mm}
\begin{equation}
\begin{tabular}{ll}
$\left\vert \varepsilon _{su\left( N\right) }\left( \mathbf{k}\right)
\right\vert =t_{1}\sqrt{N+2\dsum\limits_{i<j=0}^{N-1}\cos \left[ a\mathbf{k}%
.\left( \mathbf{\lambda }_{i}\mathbf{-\lambda }_{j}\right) \right] }$ &
\end{tabular}%
\end{equation}%
\vspace{2mm}
with $N=2,3$. However, in the case of $sp^{3}$, the atoms have no
delocalized pi-electrons; they only have strongly correlated sigma-
electrons which make the electronic properties of systems based on $\mathcal{%
L}_{su\left( 4\right) }$ different from those based on $\mathcal{L}%
_{su\left( 3\right) }$ and $\mathcal{L}_{su\left( 2\right) }$. Nevertheless,
as far as tight binding model idea is concerned, one may consider other
applications; one of which concerns the following toy model describing a
system based on the lattice $\mathcal{L}_{su\left( 4\right) }$ with
dynamical vacancy sites.
\begin{figure}[tbph]
\centering
\hspace{0cm} \includegraphics[width=5cm]{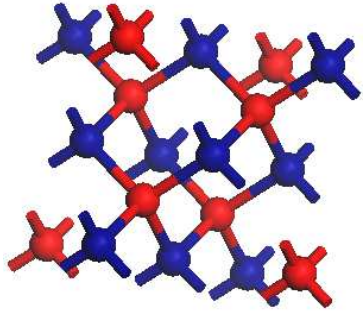}
\caption{{ the lattice }$\mathcal{L}_{su\left( 4\right) }$
with { sublattices }$\mathcal{A}_{su\left( 4\right) }$%
{ \ (in blue) and }$\mathcal{B}_{su\left( 4\right) }$%
{ \ (in red). Each atom has \emph{4} first nearest neighbors,
forming a tetrahedron, and \emph{12} second nearest ones.}}
\label{4}
\end{figure}

\emph{2)} \emph{Toy model}\newline
This is a lattice QFT on the $\mathcal{L}_{su\left( 4\right) }$ with
dynamical particles and vacancies. The initial state of the system
correspond to the configuration where the sites of the sublattice $\mathcal{A%
}_{4}$ are occupied by particles and those of the sublattice $\mathcal{B}%
_{4} $ are unoccupied.%
\vspace{5mm}

\begin{equation}
\begin{tabular}{l|l|l}
{\small sublattice} & {\small initial configuration} & {\small quantum state}
\\ \hline
$\mathcal{A}_{_{4}}$ & {\small particles at }$\mathbf{r}_{m}$ & $\mathbf{A}_{%
\mathbf{r}_{m}}$ \\
$\mathcal{B}_{_{4}}$ & {\small vacancy at }$\mathbf{r}_{m}+\mathbf{v}$ & $%
\mathbf{B}_{\mathbf{r}_{m}+\mathbf{v}}$ \\ \hline
\end{tabular}%
\end{equation}%

Then, the particles (vacancies) start to move towards the neighboring sites
with movement modeled by hops to first nearest neighbors. Let $A_{\mathbf{r}%
_{m}}$ and $B_{\mathbf{r}_{m}+\mathbf{v}_{i}}$ be the quantum states
describing the particle at $\mathbf{r}_{m}$ and the vacancy at $\mathbf{r}%
_{m}+\mathbf{v}_{i}$ respectively. Let also $A_{\mathbf{r}_{m}}^{\pm }$ and $%
B_{\mathbf{r}_{m}+\mathbf{v}_{i}}^{\pm }$ be the corresponding creation and
annihilation operators. The hamiltonian describing the hops of the
vacancy/particle to the first nearest neighbors is given by%
\begin{equation}
\begin{tabular}{ll}
$H_{_{4}}=$ & $-t\left( \dsum\limits_{i=0}^{3}A_{\mathbf{r}_{m}}^{-}B_{%
\mathbf{r}_{m}+\upsilon _{i}}^{+}+hc\right) .$%
\end{tabular}%
\end{equation}%
By performing the Fourier transform of the wave functions $A_{\mathbf{r}%
_{m}}^{\pm }$, $B_{\mathbf{r}_{m}+\upsilon _{i}}^{\pm }$, we end with the
dispersion energy $\pm t\left\vert \varepsilon _{\mathbf{k}}\right\vert $
where%
\begin{equation}
\begin{tabular}{ll}
$\varepsilon _{\mathbf{k}}=\sqrt{4+2\dsum\limits_{i<j}\cos \left( \mathbf{k.V%
}_{ij}\right) }$ & ,%
\end{tabular}%
\end{equation}%
and $\mathbf{V}_{ij}$ are as in (\ref{v0}-\ref{R}). The Dirac points are
located at $\mathbf{k}_{s}=\pm \frac{s\pi }{2a}\sum_{i=1}^{3}\mathbf{\omega }%
_{i}$ with $s=1,2$.

\section{Four dimensional graphene}

The so called four dimensional graphene is a QFT model that lives on the
\emph{4D} hyperdiamond; it has links with lattice quantum chromodynamics
(QCD) to be discussed in next section. In this section, we first study the
\emph{4D} hyperdiamond; then we use the results of previous section to give
some physical properties of \emph{4D} graphene.

\subsection{Four dimensional hyperdiamond}

\begin{figure}[tbph]
\centering
\hspace{0cm} \includegraphics[width=10cm]{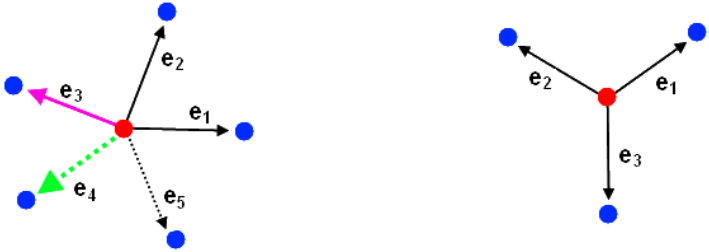}
\caption{{ On left the }\emph{5}{ \ first
nearest neighbors in the pristine }\emph{4D} { hyperdiamond
with the properties }$\left\Vert \mathbf{e}_{i}\right\Vert =1$
{ and }$e_{1}+e_{2}+e_{3}+\mathbf{e}_{4}+\mathbf{e}_{5}=0$.
{ On right, the} \emph{3} { first nearest in
pristine 2D graphene with }$\left\Vert \mathbf{e}_{i}\right\Vert =1$%
{ \ and }$e_{1}+\mathbf{e}_{2}+\mathbf{e}_{3}=0$. }
\label{5NE}
\end{figure}

Like in the case of \emph{2D} honeycomb, the \emph{4D} hyperdiamond may be
defined by the superposition of two sublattices $\mathcal{A}_{4}$ and $%
\mathcal{B}_{4}$ with the following properties:

\begin{itemize}
\item sites in $\mathcal{A}_{4}$ and $\mathcal{B}_{4}$ are parameterized by
the typical \emph{4d}- vectors $\mathbf{r}_{\mathbf{n}}$ with $\mathbf{n}%
=\left( {\small n}_{1}{\small ,n}_{2}{\small ,n}_{3}{\small ,n}_{4}\right) $
and ${\small n}_{i}$'s arbitrary integers. These lattice vectors are
expanded as follows
\begin{equation}
\begin{tabular}{llll}
$\mathcal{A}_{4}$: $\ \mathbf{r}_{\mathbf{n}}=n_{1}$ $\mathbf{a}_{1}+n_{2}$ $%
\mathbf{a}_{2}+n_{3}$ $\mathbf{a}_{3}+n_{4}$ $\mathbf{a}_{4}$ & $,$ & $%
\mathcal{B}_{4}$: $\ \mathbf{r}_{\mathbf{n}}^{\prime }=\mathbf{r}_{\mathbf{n}%
}+\mathbf{e}_{5}$ & $,$%
\end{tabular}
\label{Z1}
\end{equation}%
where $\mathbf{a}_{1},$ $\mathbf{a}_{2},$ $\mathbf{a}_{3},$ $\mathbf{a}_{4}$
are primitive vectors generating these sublattices; and $\mathbf{e}_{5}$ is
a shift vector which we describe below.

\item the vector $\mathbf{e}_{5}$ is a global vector taking the same value $%
\forall $ $\mathbf{n}$; it is a shift vector giving the relative positions
of the $\mathcal{B}_{4}$ sites with respect to the $\mathcal{A}_{4}$ ones, $%
\mathbf{e}_{5}\mathbf{=r}_{\mathbf{n}}^{\prime }-\mathbf{r}_{\mathbf{n}}$, $%
\forall $ $\mathbf{n}$.
\end{itemize}

\ \ \ \newline
The $\mathbf{a}_{l}$'s and $\mathbf{e}_{5}$ vectors can be chosen as
\begin{equation}
\begin{tabular}{lllll}
$\mathbf{a}_{1}=\mathbf{e}_{1}-\mathbf{e}_{5},$ & $\mathbf{a}_{2}=\mathbf{e}%
_{2}-\mathbf{e}_{5},$ & $\mathbf{a}_{3}=\mathbf{e}_{3}-\mathbf{e}_{5}$, & $%
\mathbf{a}_{4}=\mathbf{e}_{4}-\mathbf{e}_{5}$ &
\end{tabular}%
\end{equation}%
with
\begin{equation}
\begin{tabular}{lll}
$\mathbf{e}_{1}^{\mu }$ & $=\frac{1}{4}\left( +\sqrt{5},+\sqrt{5},+\sqrt{5}%
,+1\right) $ & , \\
$\mathbf{e}_{2}^{\mu }$ & $=\frac{1}{4}\left( +\sqrt{5},-\sqrt{5},-\sqrt{5}%
,+1\right) $ & , \\
$\mathbf{e}_{3}^{\mu }$ & $=\frac{1}{4}\left( -\sqrt{5},-\sqrt{5},+\sqrt{5}%
,+1\right) $ & , \\
$\mathbf{e}_{4}^{\mu }$ & $=\frac{1}{4}\left( -\sqrt{5},+\sqrt{5},-\sqrt{5}%
,+1\right) $ & ,%
\end{tabular}
\label{Z2}
\end{equation}%
and $\mathbf{e}_{5}=-\sum_{i=1}^{4}\mathbf{e}_{i}$. Notice also that the
\emph{5} vectors $\mathbf{e}_{1},$ $\mathbf{e}_{2},$ $\mathbf{e}_{3},$ $%
\mathbf{e}_{4},$ $\mathbf{e}_{5}$ define the first nearest neighbors to $%
\left( 0,0,0,0\right) $ and satisfy the constraint relations,
\begin{equation}
\begin{tabular}{lll}
$\mathbf{e}_{i}.\mathbf{e}_{i}=1,$ & $\mathbf{e}_{i}.\mathbf{e}_{j}=\cos
\vartheta _{ij}=-\frac{1}{4},\quad i\neq j$ & ,%
\end{tabular}
\label{Z3}
\end{equation}%
showing that the $\mathbf{e}_{i}$'s are distributed in a symmetric way since
all the angles satisfy $\cos \vartheta _{ij}=\frac{-1}{4}$; see also figure (%
\ref{5NE}) for illustration.

\textbf{some specific properties}\newline
From the figure (\ref{5NE}) representing the first nearest neighbors in the
\emph{4D} hyperdiamond and their analog in \emph{2D} graphene, we learn that
each $\mathcal{A}_{4}$- type node at $\mathbf{r}_{\mathbf{n}}$, with some
attached wave function $A_{\mathbf{r}_{\mathbf{n}}}$, has the following
closed neighbors:

\begin{itemize}
\item \emph{5} first nearest neighbors belonging to $\mathcal{B}_{4}$ with
wave functions $B_{\mathbf{r}_{\mathbf{n}}+d\mathbf{e}_{i}}$; they are given
by:%
\begin{equation}
\begin{tabular}{l|l}
\textit{lattice position} & \textit{attached wave} \\ \hline
$\ \ \ \mathbf{r}_{\mathbf{n}}+d\mathbf{e}_{1}$ & $\ \ \ B_{\mathbf{r}_{%
\mathbf{n}}+d\mathbf{e}_{1}}$ \\
$\ \ \ \mathbf{r}_{\mathbf{n}}+d\mathbf{e}_{2}$ & $\ \ \ B_{\mathbf{r}_{%
\mathbf{n}}+d\mathbf{e}_{2}}$ \\
$\ \ \ \mathbf{r}_{\mathbf{n}}+d\mathbf{e}_{3}$ & $\ \ \ B_{\mathbf{r}_{%
\mathbf{n}}+d\mathbf{e}_{3}}$ \\
$\ \ \ \mathbf{r}_{\mathbf{n}}+d\mathbf{e}_{4}$ & $\ \ \ B_{\mathbf{r}_{%
\mathbf{n}}+d\mathbf{e}_{4}}$ \\
$\ \ \ \mathbf{r}_{\mathbf{n}}+d\mathbf{e}_{5}$ & $\ \ \ B_{\mathbf{r}_{%
\mathbf{n}}+d\mathbf{e}_{5}}$ \\ \hline
\end{tabular}%
\end{equation}%
Using this configuration, the typical tight binding hamiltonian describing
the couplings between the first nearest neighbors reads as
\begin{equation}
\begin{tabular}{ll}
$-t\dsum\limits_{\mathbf{r}_{\mathbf{n}}}\dsum\limits_{i=1}^{5}A_{\mathbf{r}%
_{\mathbf{n}}}B_{\mathbf{r}_{\mathbf{n}}+d\mathbf{e}_{i}}^{+}+hc$ & .%
\end{tabular}%
\end{equation}%
where $t$ is the hop energy and where $d$ is the lattice parameter.\newline
Notice that in the case where the wave functions at $\mathbf{r}_{\mathbf{n}}$
and $\mathbf{r}_{\mathbf{n}}+d\mathbf{e}_{i}$ are rather given by two
component Weyl spinors
\begin{equation}
\begin{tabular}{llll}
$A_{\mathbf{r}_{\mathbf{n}}}^{a}=\left(
\begin{array}{c}
A_{\mathbf{r}_{\mathbf{n}}}^{1} \\
A_{\mathbf{r}_{\mathbf{n}}}^{2}%
\end{array}%
\right) $ & , & $\bar{B}_{\mathbf{r}_{\mathbf{n}}+d\mathbf{e}_{i}}^{\dot{a}%
}=\left(
\begin{array}{c}
\bar{B}_{\mathbf{r}_{\mathbf{n}}+d\mathbf{e}_{i}}^{\dot{1}} \\
\bar{B}_{\mathbf{r}_{\mathbf{n}}+d\mathbf{e}_{i}}^{\dot{2}}%
\end{array}%
\right) $ & ,%
\end{tabular}%
\end{equation}%
together with their adjoints $\bar{A}_{\mathbf{r}_{\mathbf{n}}}^{\dot{a}}$
and $\bar{B}_{\mathbf{r}_{\mathbf{n}}+d\mathbf{e}_{i}}^{a}$, as in the
example of \emph{4D} lattice QCD to be described in next section, the
corresponding tight binding model would be,
\begin{equation}
\begin{tabular}{ll}
$-t\dsum\limits_{\mathbf{r}_{\mathbf{n}}}\dsum\limits_{i=1}^{5}\left[
\dsum\limits_{\mu =1}^{4}\mathbf{e}_{i}^{\mu }\left( A_{\mathbf{r}_{\mathbf{n%
}}}^{a}\sigma _{a\dot{a}}^{\mu }\bar{B}_{\mathbf{r}_{\mathbf{n}}+d\mathbf{e}%
_{i}}^{\dot{a}}\right) \right] +hc$ & .%
\end{tabular}%
\end{equation}%
where the $\mathbf{e}_{i}^{\mu }$'s are as in (\ref{Z2}); and where $\sigma
^{1},$ $\sigma ^{2},$ $\sigma ^{3}$ are the Pauli matrices and $\sigma
^{4}=I_{2\times 2}$. Notice moreover that the term $\sum_{i=1}^{5}\mathbf{e}%
_{i}^{\mu }\left( A_{\mathbf{r}_{\mathbf{n}}}^{a}\sigma _{a\dot{a}}^{\mu }%
\bar{B}_{\mathbf{r}_{\mathbf{n}}}^{\dot{a}}\right) $ vanishes identically
due to $\sum_{i=1}^{5}\mathbf{e}_{i}^{\mu }=0.$

\item \emph{20} second nearest neighbors belonging to the same $\mathcal{A}%
_{4}$ with the wave functions $A_{\mathbf{r}_{\mathbf{n}}+d\left( \mathbf{e}%
_{i}-\mathbf{e}_{j}\right) }$; they read as%
\begin{equation}
\begin{tabular}{llll}
${\small r}_{\mathbf{n}}\pm d\left( \mathbf{e}_{1}-\mathbf{e}_{2}\right)
{\small ,}$ & ${\small r}_{\mathbf{n}}\pm d\left( \mathbf{e}_{1}-\mathbf{e}%
_{3}\right) {\small ,}$ & ${\small r}_{\mathbf{n}}\pm d\left( \mathbf{e}_{1}-%
\mathbf{e}_{4}\right) {\small ,}$ &  \\
${\small r}_{\mathbf{n}}\pm d\left( \mathbf{e}_{1}-\mathbf{e}_{5}\right)
{\small ,}$ & ${\small r}_{\mathbf{n}}\pm d\left( \mathbf{e}_{2}-\mathbf{e}%
_{3}\right) {\small ,}$ & ${\small r}_{\mathbf{n}}\pm d\left( \mathbf{e}_{2}-%
\mathbf{e}_{4}\right) {\small ,}$ &  \\
${\small r}_{\mathbf{n}}\pm d\left( \mathbf{e}_{2}-\mathbf{e}_{5}\right)
{\small ,}$ & ${\small r}_{\mathbf{n}}\pm d\left( \mathbf{e}_{3}-\mathbf{e}%
_{4}\right) {\small ,}$ & ${\small r}_{\mathbf{n}}\pm d\left( \mathbf{e}_{3}-%
\mathbf{e}_{5}\right) {\small ,}$ &  \\
${\small r}_{\mathbf{n}}\pm d\left( \mathbf{e}_{4}-\mathbf{e}_{5}\right)
{\small .}$ &  &  &
\end{tabular}
\label{Z4}
\end{equation}
\end{itemize}

\ \ \ \newline
The \emph{5} vectors $\mathbf{e}_{1},$ $\mathbf{e}_{2},$ $\mathbf{e}_{3},$ $%
\mathbf{e}_{4},$ $\mathbf{e}_{5}$ are, up to a normalization factor namely $%
\frac{\sqrt{5}}{2}$, precisely the weight vectors $\mathbf{\lambda }_{0},$ $%
\mathbf{\lambda }_{1},$ $\mathbf{\lambda }_{2},$ $\mathbf{\lambda }_{3},$ $%
\mathbf{\lambda }_{4}$ of the \emph{5}-dimensional representation of $%
SU\left( 5\right) $; and the \emph{20} vectors $\left( \mathbf{e}_{i}-%
\mathbf{e}_{j}\right) $ are, up to a scale factor $\frac{\sqrt{5}}{2}$,
their roots $\beta _{ij}=\left( \mathbf{\lambda }_{i}-\mathbf{\lambda }%
_{j}\right) $. We show as well that the particular property $\mathbf{e}_{i}.%
\mathbf{e}_{j}=-\frac{1}{4}$, which is constant $\forall $ $\mathbf{e}_{i}$
and $\mathbf{e}_{j}$, has a natural interpretation in terms of the Cartan
matrix of $SU\left( 5\right) $.

\textbf{2D/4D Correspondence}\emph{\newline
}First\emph{\ }notice that a generic bond vector $\mathbf{e}_{i}$ in the
hyperdiamond links two sites in the same unit cell of the \emph{4D} lattice
as shown on the typical coupling term $A_{\mathbf{r}_{\mathbf{n}}}B_{\mathbf{%
r}_{\mathbf{n}}+d\mathbf{e}_{i}}^{+}$. This property is quite similar to the
action of the usual $\gamma ^{\mu }$ matrices on \emph{4D} (Euclidean) space
time spinors which links the components of spinors.\newline
Mimicking the tight binding model of \emph{2D} graphene, it has been
proposed in \textrm{\citep{E2}} a graphene inspired model for \emph{4D}
lattice QCD. There, the construction relies on the use of the following:

\begin{itemize}
\item the naive correspondence between the bond vectors $\mathbf{e}_{i}$ and
the $\mathrm{\gamma }^{i}$ matrices%
\begin{equation}
\begin{tabular}{llllll}
$\mathbf{e}_{i}$ & $\longleftrightarrow $ & $\gamma _{i}$ & , & $i=1,...,4$
& ,%
\end{tabular}%
\end{equation}%
together with%
\begin{equation}
\begin{tabular}{ll}
$-\mathbf{e}_{5}=\mathbf{e}_{1}+\mathbf{e}_{2}+\mathbf{e}_{3}+\mathbf{e}_{4}$
& , \\
$-\Gamma _{5}=\gamma _{1}+\gamma _{2}+\gamma _{3}+\gamma _{4}$ & .%
\end{tabular}%
\end{equation}

\item as in the case of \emph{2D} graphene, $\mathcal{A}_{4}$-type sites are
occupied by left $\phi _{\mathbf{r}}^{a}$ and right $\bar{\phi}_{\mathbf{r}%
}^{\dot{a}}$ 2-component Weyl spinors. $\mathcal{B}_{4}$-type sites are
occupied by right $\bar{\chi}_{\mathbf{r}+d\mathbf{e}_{i}}^{\dot{a}}$ and
left $\chi _{\mathbf{r}+d\mathbf{e}_{i}}^{a}$ Weyl spinors.
\begin{equation}
\begin{tabular}{|l|l|l|}
\hline
\ \ \ \ lattice & {\small 2D graphene} & {\small 4D hyperdiamond} \\ \hline
$\mathcal{A}_{4}$-sites at \ $\mathbf{r}_{n}$ & $A_{\mathbf{r}}$ &
\begin{tabular}{ll}
$\phi _{\mathbf{r}}^{a},$ & $\bar{\phi}_{\mathbf{r}}^{\dot{a}}$%
\end{tabular}
\\ \hline
$\mathcal{B}_{4}$-sites at \ $\mathbf{r}_{n}+d\mathbf{e}_{i}$ & $B_{\mathbf{r%
}+de_{i}}^{+}$ &
\begin{tabular}{ll}
$\bar{\chi}_{\mathbf{r}+d\mathbf{e}_{i}}^{\dot{a}},$ & $\chi _{\mathbf{r}+d%
\mathbf{e}_{i}}^{a}$%
\end{tabular}
\\ \hline
\ \ couplings & $%
\begin{array}{c}
A_{\mathbf{r}}B_{\mathbf{r}+de_{i}}^{+} \\
B_{\mathbf{r}+de_{i}}A_{\mathbf{r}}^{+}%
\end{array}%
$ & $%
\begin{array}{c}
\mathbf{e}_{i}^{\mu }\left( \phi _{\mathbf{r}}^{a}\sigma _{a\dot{a}}^{\mu }%
\text{ }\bar{\chi}_{\mathbf{r}+d\mathbf{e}_{i}}^{\dot{a}}\right) \\
\mathbf{e}_{i}^{\mu }\left( \chi _{\mathbf{r}+d\mathbf{e}_{i}}^{a}\bar{\sigma%
}_{a\dot{a}}^{\mu }\text{ }\bar{\phi}_{\mathbf{r}}^{\dot{a}}\right)%
\end{array}%
$ \\ \hline
\end{tabular}%
\end{equation}%
where the indices $a=1,2$ and $\dot{a}=\dot{1},\dot{2}$; and where summation
over $\mu $ is in the Euclidean sense.
\end{itemize}

\ \ \ \ \newline
For later use, it is interesting to notice the two following:

\begin{description}
\item[(a)] in \emph{2D} graphene, the wave functions $A_{\mathbf{r}}$ and $%
B_{\mathbf{r}+de_{i}}$ describe polarized electrons in first nearest sites
of the \emph{2D} honeycomb. As the spin up and spin down components of the
electrons contribute equally, the effect of spin couplings in \emph{2D}
graphene is ignored.

\item[(b)] in the \emph{4D} hyperdiamond, we have \emph{4+4} wave functions
at each $\mathcal{A}_{4}$-type site or $\mathcal{B}_{4}$-type one. These
wave functions are given by:

\begin{description}
\item[(i)] $\phi ^{a}=\left( \phi _{\mathbf{r}_{n}}^{1},\phi _{\mathbf{r}%
_{n}}^{2}\right) $ and $\bar{\phi}_{\mathbf{r}}^{\dot{a}}=\left( \bar{\phi}_{%
\mathbf{r}_{n}}^{\dot{1}},\bar{\phi}_{\mathbf{r}_{n}}^{\dot{2}}\right) $
having respectively positive and negative $\mathrm{\gamma }^{5}$ chirality,

\item[(ii)] $\bar{\chi}_{\mathbf{r}+d\mathbf{e}_{i}}^{\dot{a}}=\left( \bar{%
\chi}_{\mathbf{r}+d\mathbf{e}_{i}}^{\dot{1}},\bar{\chi}_{\mathbf{r}+d\mathbf{%
e}_{i}}^{\dot{2}}\right) $ and $\chi _{\mathbf{r}+d\mathbf{e}%
_{i}}^{a}=\left( \chi _{\mathbf{r}+d\mathbf{e}_{i}}^{1},\chi _{\mathbf{r}+d%
\mathbf{e}_{i}}^{2}\right) $ having respectively negative and positive $%
\gamma ^{5}$ chirality.
\end{description}
\end{description}

By mimicking the \emph{2D }graphene study, we have the couplings%
\begin{equation}
\begin{tabular}{llll}
$\mathbf{e}_{i}^{\mu }\sigma _{1\dot{1}}^{\mu }\left( \phi _{\mathbf{r}}^{1}%
\text{ }\bar{\chi}_{\mathbf{r}+d\mathbf{e}_{i}}^{\dot{1}}\right) $ & , & $%
\mathbf{e}_{i}^{\mu }\sigma _{2\dot{2}}^{\mu }\left( \phi _{\mathbf{r}}^{2}%
\text{ }\bar{\chi}_{\mathbf{r}+d\mathbf{e}_{i}}^{\dot{2}}\right) $ &  \\
$\mathbf{e}_{i}^{\mu }\bar{\sigma}_{1\dot{1}}^{\mu }\left( \chi _{\mathbf{r}%
+d\mathbf{e}_{i}}^{1}\text{ }\bar{\phi}_{\mathbf{r}}^{\dot{1}}\right) $ & ,
& $\mathbf{e}_{i}^{\mu }\bar{\sigma}_{2\dot{2}}^{\mu }\left( \chi _{\mathbf{r%
}+d\mathbf{e}_{i}}^{2}\text{ }\bar{\phi}_{\mathbf{r}}^{\dot{2}}\right) $ &
\end{tabular}%
\end{equation}

\textbf{building the hamiltonian}\newline
To describe 4D lattice fermions, one considers \emph{4D} space time Dirac
spinors together with the following $\gamma ^{\mu }$ matrices realizations,%
\begin{equation}
\begin{tabular}{llllll}
$\gamma ^{1}=\tau ^{1}\otimes \sigma ^{1}$ & , & $\gamma ^{2}=\tau
^{1}\otimes \sigma ^{2}$ & , & $\gamma ^{3}=\tau ^{1}\otimes \sigma ^{3}$ & ,
\\
$\gamma ^{4}=\tau ^{2}\otimes I_{2}$ & , & $\gamma ^{5}=\tau ^{3}\otimes
I_{2}$ & , &  &
\end{tabular}
\label{515}
\end{equation}%
where the $\tau ^{i}$'s are the Pauli matrices acting on the sublattice
structure of the hyperdiamond lattice,
\begin{equation}
\begin{tabular}{llll}
$\tau ^{1}=\left(
\begin{array}{cc}
0 & 1 \\
1 & 0%
\end{array}%
\right) ,$ & $\tau ^{2}=\left(
\begin{array}{cc}
0 & -i \\
i & 0%
\end{array}%
\right) ,$ & $\tau ^{3}=\left(
\begin{array}{cc}
1 & 0 \\
0 & -1%
\end{array}%
\right) .$ &
\end{tabular}%
\end{equation}%
The $2\times 2$ matrices $\sigma ^{i}$ satisfy as well the Clifford algebra $%
\sigma ^{i}\sigma ^{j}+\sigma ^{j}\sigma ^{i}=2\delta ^{ij}I_{2}$ and act
through the coupling of left/right Weyl spinors at neighboring sites%
\begin{equation}
\phi _{\mathbf{r}}^{a}\sigma _{a\dot{a}}^{\mu }\text{ }\bar{\chi}_{\mathbf{r}%
+d\frac{\sqrt{5}}{2}\mathbf{\lambda }_{i}}^{\dot{a}}-\chi _{\mathbf{r}}^{a}%
\bar{\sigma}_{a\dot{a}}^{\mu }\text{ }\bar{\phi}_{\mathbf{r-}d\frac{\sqrt{5}%
}{2}\mathbf{\lambda }_{i}}^{\dot{a}}=\left( \phi _{\mathbf{r}}\sigma ^{\mu }%
\bar{\chi}_{\mathbf{r}+d\frac{\sqrt{5}}{2}\mathbf{\lambda }_{i}}-\chi _{%
\mathbf{r}}\bar{\sigma}^{\mu }\bar{\phi}_{\mathbf{r}-d\frac{\sqrt{5}}{2}%
\mathbf{\lambda }_{i}}\right)
\end{equation}%
where $\sigma ^{\mu }=\left( \sigma ^{1},\sigma ^{2},\sigma
^{3},+iI_{2}\right) $ and $\bar{\sigma}^{\mu }=\left( \sigma ^{1},\sigma
^{2},\sigma ^{3},-iI_{2}\right) .$ For later use, it is interesting to set%
\begin{equation}
\begin{tabular}{lll}
$\sigma ^{\mu }.e_{1}^{\mu }$ & $=\frac{\sqrt{5}}{4}\sigma ^{1}+\frac{\sqrt{5%
}}{4}\sigma ^{2}+\frac{\sqrt{5}}{4}\sigma ^{3}+\frac{i}{4}I_{2}$, &  \\
$\bar{\sigma}^{\mu }.e_{1}^{\mu }$ & $=\frac{\sqrt{5}}{4}\sigma ^{1}+\frac{%
\sqrt{5}}{4}\sigma ^{2}+\frac{\sqrt{5}}{4}\sigma ^{3}-\frac{i}{4}I_{2}$, &
\end{tabular}%
\end{equation}%
and similar relations for the other $\sigma .e_{i}$ and $\bar{\sigma}\mathrm{%
.}\mathbf{e}_{i}$.

\ \ \ \ \newline
Now extending the tight binding model of 2D graphene to the 4D hyperdiamond;
and using the weight vectors $\lambda _{i}$ instead of $e_{i}$, we can build
a free fermion action on the 4D lattice by attaching a two-component
left-handed spinor $\phi ^{a}\left( \mathbf{r}\right) $ and right-handed
spinor $\bar{\phi}_{\mathbf{r}}^{\dot{a}}$ to each $A_{4}$-node $r$, and a
right-handed spinor $\bar{\chi}_{\mathbf{r}+d\frac{\sqrt{5}}{2}\mathbf{%
\lambda }_{i}}^{\dot{a}}$ and left-handed spinor $\chi _{\mathbf{r}+d\frac{%
\sqrt{5}}{2}\mathbf{\lambda }_{i}}^{a}$ to every $B_{4}$-node at $r+d\frac{%
\sqrt{5}}{2}\lambda _{i}$. \newline
The hamiltonian, describing hopping to first nearest-neighbor sites with
equal probabilities in all five directions $\mathbf{\lambda }_{i}$, reads as
follows:.%
\begin{equation}
\begin{tabular}{llll}
$H_{4}$ & $=$ & $\dsum\limits_{\mathbf{r}}\dsum\limits_{i=0}^{4}\left( \phi
_{\mathbf{r}}\sigma ^{\mu }\bar{\chi}_{\mathbf{r+}d\frac{\sqrt{5}}{2}\mathbf{%
\lambda }_{i}}-\chi _{\mathbf{r}}\bar{\sigma}^{\mu }\bar{\phi}_{\mathbf{r-}d%
\frac{\sqrt{5}}{2}\mathbf{\lambda }_{i}}\right) \lambda _{i}^{\mu }$ & .%
\end{tabular}
\label{AC}
\end{equation}%
Expanding the various spinorial fields $\xi _{\mathbf{r\pm v}}$ in Fourier
sums as $\int \frac{d^{4}k}{\left( 2\pi \right) ^{4}}e^{-i\mathbf{k.r}%
}\left( e^{\mp i\mathbf{k.v}}\xi _{\mathbf{k}}\right) $ with $\mathbf{k}$
standing for a generic wave vector in the reciprocal lattice, we can put the
field action $H_{4}$ into the form
\begin{equation}
\begin{tabular}{llll}
$H_{4}$ & $=$ & $i\dsum\limits_{\mathbf{k}}\left( \bar{\phi}_{\mathbf{k}},%
\bar{\chi}_{\mathbf{k}}\right) \left(
\begin{array}{cc}
0 & -iD \\
i\bar{D} & 0%
\end{array}%
\right) \left(
\begin{array}{c}
\phi _{\mathbf{k}} \\
\chi _{\mathbf{k}}%
\end{array}%
\right) $ &
\end{tabular}
\label{D1}
\end{equation}%
where we have set%
\begin{equation}
\begin{tabular}{llll}
$D$ & $=\dsum\limits_{l=0}^{4}D_{l}e^{id\frac{\sqrt{5}}{2}\mathbf{k}.\lambda
_{l}}$ & $=\dsum\limits_{\mu =1}^{4}\sigma ^{\mu }\left(
\dsum\limits_{l=0}^{4}\lambda _{l}^{\mu }e^{id\frac{\sqrt{5}}{2}\mathbf{k}%
.\lambda _{l}}\right) $ & ,%
\end{tabular}
\label{D2}
\end{equation}%
with%
\begin{equation}
\begin{tabular}{ll}
$D_{l}=\dsum\limits_{\mu =1}^{4}\sigma ^{\mu }\lambda _{l}^{\mu }=\left(
\begin{array}{cc}
\lambda _{l}^{3}+i\lambda _{l}^{4} & \lambda _{l}^{1}-i\lambda _{l}^{2} \\
\lambda _{l}^{1}+i\lambda _{l}^{2} & \lambda _{l}^{3}-i\lambda _{l}^{4}%
\end{array}%
\right) $ & ,%
\end{tabular}%
\end{equation}%
and $p_{l}=\mathbf{k}.\lambda _{l}=\sum_{\mu }k_{\mu }\lambda _{l}^{\mu }$.
Similarly we have%
\begin{equation}
\begin{tabular}{llll}
$\bar{D}$ & $=\dsum\limits_{l=0}^{4}\bar{D}_{l}e^{-id\frac{\sqrt{5}}{2}%
\mathbf{k}.\lambda _{l}}$ & $=\dsum\limits_{\mu =1}^{4}\bar{\sigma}^{\mu
}\left( \dsum\limits_{l=0}^{4}\lambda _{l}^{\mu }e^{-id\frac{\sqrt{5}}{2}%
\mathbf{k}.\lambda _{l}}\right) $ & .%
\end{tabular}
\label{2D}
\end{equation}

\subsection{Energy dispersion and zero modes}

To get the dispersion energy relations of the 4 waves components $\phi _{%
\mathbf{k}}^{1}$, $\phi _{\mathbf{k}}^{2}$, $\chi _{\mathbf{k}}^{1},$ $\chi
_{\mathbf{k}}^{2}$ and their corresponding 4 holes, one has to solve the
eigenvalues of the Dirac operator (\ref{D1}). To that purpose, we first
write the 4-dimensional wave equation as follows,%
\begin{equation}
\begin{tabular}{llll}
$\left(
\begin{array}{cc}
0 & -iD \\
i\bar{D} & 0%
\end{array}%
\right) \left(
\begin{array}{c}
\phi _{\mathbf{k}} \\
\chi _{\mathbf{k}}%
\end{array}%
\right) $ & $=$ & $E\left(
\begin{array}{c}
\phi _{\mathbf{k}} \\
\chi _{\mathbf{k}}%
\end{array}%
\right) $ & ,%
\end{tabular}
\label{DE}
\end{equation}%
where $\phi _{\mathbf{k}}=\left( \phi _{\mathbf{k}}^{1},\phi _{\mathbf{k}%
}^{2}\right) $, $\chi _{\mathbf{k}}=\left( \chi _{\mathbf{k}}^{1},\chi _{%
\mathbf{k}}^{2}\right) $ are Weyl spinors and where the $2\times 2$ matrices
$D$, $\bar{D}$ are as in eqs(\ref{D2},\ref{2D}). Then determine the
eigenstates and eigenvalues of the $2\times 2$ Dirac operator matrix by
solving the following characteristic equation,%
\begin{equation}
\begin{tabular}{lll}
$\det \left(
\begin{array}{cccc}
-E & 0 & D_{11} & D_{12} \\
0 & -E & D_{21} & D_{22} \\
\bar{D}_{11} & \bar{D}_{21} & -E & 0 \\
\bar{D}_{12} & \bar{D}_{22} & 0 & -E%
\end{array}%
\right) $ & $=0$ &
\end{tabular}%
\end{equation}%
from which one can learn the four dispersion energy eigenvalues $E_{1}\left(
\mathbf{k}\right) $, $E_{2}\left( \mathbf{k}\right) $, $E_{3}\left( \mathbf{k%
}\right) $, $E_{4}\left( \mathbf{k}\right) $ and therefore their zeros.

\textbf{1) computing the energy dispersion}\newline
An interesting way to do these calculations is to act on (\ref{DE}) once
more by the Dirac operator to bring it to the following diagonal form%
\begin{equation}
\begin{tabular}{llll}
$\left(
\begin{array}{cc}
D\bar{D} & 0 \\
0 & D\bar{D}%
\end{array}%
\right) \left(
\begin{array}{c}
\phi _{\mathbf{k}} \\
\chi _{\mathbf{k}}%
\end{array}%
\right) $ & $=$ & $E^{2}\left(
\begin{array}{c}
\phi _{\mathbf{k}} \\
\chi _{\mathbf{k}}%
\end{array}%
\right) $ & .%
\end{tabular}%
\end{equation}%
Then solve separately the eigenvalues problem of the 2-dimensional equations
$D\bar{D}\phi _{\mathbf{k}}=E^{2}\phi _{\mathbf{k}}$ and $\bar{D}D\chi _{%
\mathbf{k}}=E^{2}\chi _{\mathbf{k}}$. To do so, it is useful to set
\begin{equation}
\begin{tabular}{llll}
$u\left( \mathbf{k}\right) =\vartheta ^{1}+i\vartheta ^{2}$ & , & $v\left(
\mathbf{k}\right) =\vartheta ^{3}+i\vartheta ^{4}$ &
\end{tabular}
\label{uv}
\end{equation}%
with $\vartheta ^{\mu }=\sum_{l}\lambda _{l}^{\mu }\exp (id\frac{\sqrt{5}}{2}%
k.\lambda _{l})$. Notice that in the continuous limit, we have $\vartheta
^{\mu }\rightarrow id\frac{\sqrt{5}}{2}\mathbf{k}^{\mu }$,%
\begin{equation}
\begin{tabular}{lll}
$u\left( \mathbf{k}\right) \rightarrow id\frac{\sqrt{5}}{2}\left( \mathbf{k}%
^{1}+i\mathbf{k}^{2}\right) $, & $v\left( \mathbf{k}\right) \rightarrow id%
\frac{\sqrt{5}}{2}\left( \mathbf{k}^{3}+i\mathbf{k}^{4}\right) $ & .%
\end{tabular}%
\end{equation}%
Substituting (\ref{uv}) back into (\ref{D2}) and (\ref{2D}), we obtain the
following expressions,%
\begin{equation}
\begin{tabular}{lll}
${\small D\bar{D}=}\left(
\begin{array}{cc}
\left\vert u\right\vert ^{2}{\small +}\left\vert v\right\vert ^{2} & {\small %
2\bar{u}v} \\
{\small 2u\bar{v}} & \left\vert u\right\vert ^{2}{\small +}\left\vert
v\right\vert ^{2}%
\end{array}%
\right) ,$ & ${\small \bar{D}D=}\left(
\begin{array}{cc}
\left\vert u\right\vert ^{2}{\small +}\left\vert v\right\vert ^{2} & {\small %
2\bar{u}\bar{v}} \\
{\small 2uv} & \left\vert u\right\vert ^{2}{\small +}\left\vert v\right\vert
^{2}%
\end{array}%
\right) $ &
\end{tabular}%
\end{equation}%
By solving the characteristic equations of these $2\times 2$ matrix
operators, we get the eigenstates $\phi _{\mathbf{k}}^{a\prime }$, $\chi _{%
\mathbf{k}}^{a\prime }$ with their corresponding eigenvalues $E_{\pm }^{2}$,%
\begin{equation}
\begin{tabular}{l|l|l}
\multicolumn{2}{l|}{\small \ \ \ \ \ \ \ \ \ eigenstates} & {\small \ \ \ \
\ \ \ \ \ eigenvalues} \\ \hline
$\phi _{\mathbf{k}}^{1\prime }=$ & $\sqrt{\frac{v\bar{u}}{2\left\vert
u\right\vert \left\vert v\right\vert }}\phi _{\mathbf{k}}^{1}+\sqrt{\frac{u%
\bar{v}}{2\left\vert u\right\vert \left\vert v\right\vert }}\phi _{\mathbf{k}%
}^{2}$ & $E_{+}^{2}=\left\vert u\right\vert ^{2}+\left\vert v\right\vert
^{2}+2\left\vert u\right\vert \left\vert v\right\vert $ \\
$\phi _{\mathbf{k}}^{2\prime }=$ & $-\sqrt{\frac{v\bar{u}}{2\left\vert
u\right\vert \left\vert v\right\vert }}\phi _{\mathbf{k}}^{1}+\sqrt{\frac{u%
\bar{v}}{2\left\vert u\right\vert \left\vert v\right\vert }}\phi _{\mathbf{k}%
}^{2}$ & $E_{-}^{2}=\left\vert u\right\vert ^{2}+\left\vert v\right\vert
^{2}-2\left\vert u\right\vert \left\vert v\right\vert $ \\
$\chi _{\mathbf{k}}^{1\prime }=$ & $\sqrt{\frac{\bar{u}\bar{v}}{2\left\vert
u\right\vert \left\vert v\right\vert }}\chi _{\mathbf{k}}^{1}+\sqrt{\frac{uv%
}{2\left\vert u\right\vert \left\vert v\right\vert }}\chi _{\mathbf{k}}^{2}$
& $E_{+}^{2}=\left\vert u\right\vert ^{2}+\left\vert v\right\vert
^{2}+2\left\vert u\right\vert \left\vert v\right\vert $ \\
$\chi _{\mathbf{k}}^{2\prime }=$ & $-\sqrt{\frac{\bar{u}\bar{v}}{2\left\vert
u\right\vert \left\vert v\right\vert }}\chi _{\mathbf{k}}^{1}+\sqrt{\frac{uv%
}{2\left\vert u\right\vert \left\vert v\right\vert }}\chi _{\mathbf{k}}^{2}$
& $E_{-}^{2}=\left\vert u\right\vert ^{2}+\left\vert v\right\vert
^{2}-2\left\vert u\right\vert \left\vert v\right\vert $ \\ \hline
\end{tabular}%
\end{equation}%
By taking square roots of $E_{\pm }^{2}$, we obtain \emph{2} positive and
\emph{2} negative dispersion energies; these are
\begin{equation}
\begin{tabular}{ll}
$E_{\pm }=+\sqrt{\left( \left\vert u\right\vert \pm \left\vert v\right\vert
\right) ^{2}},$ & $E_{\pm }^{\ast }=-\sqrt{\left( \left\vert u\right\vert
\pm \left\vert v\right\vert \right) ^{2}}$%
\end{tabular}%
\end{equation}%
which correspond respectively to particles and the associated holes.

\textbf{2) determining the zeros of }$E_{\pm }$\textbf{\ and }$E_{\pm
}^{\ast }$\newline
From the above energy dispersion relations, one sees that the zero modes are
of two kinds: $E_{+}^{2}=0,$\emph{\ }$E_{-}^{2}=0$; and $E_{-}^{2}=0$\emph{\
but }$E_{+}^{2}=E_{+\min }^{2}\neq 0$. Let us consider the case\emph{\ }$%
E_{+}^{2}=E_{-}^{2}=0$; in this situation the zero modes are given by those
wave vectors $\mathbf{K}_{F}$ solving the constraint relations $u\left(
\mathbf{K}_{F}\right) =v\left( \mathbf{K}_{F}\right) =0$. These constraints
can be also put in the form%
\begin{equation}
\begin{tabular}{lll}
$\lambda _{0}^{\mu }e^{id\frac{\sqrt{5}}{2}\mathbf{K}_{F}.\mathrm{\lambda }%
_{0}}+\lambda _{1}^{\mu }e^{id\frac{\sqrt{5}}{2}\mathbf{K}_{F}.\mathrm{%
\lambda }_{1}}+$ &  &  \\
$+\lambda _{2}^{\mu }e^{id\frac{\sqrt{5}}{2}\mathbf{K}_{F}.\mathrm{\lambda }%
_{2}}+\lambda _{3}^{\mu }e^{id\frac{\sqrt{5}}{2}\mathbf{K}_{F}.\mathrm{%
\lambda }_{3}}+\lambda _{4}^{\mu }e^{id\frac{\sqrt{5}}{2}\mathbf{K}_{F}.%
\mathrm{\lambda }_{4}}$ & $=0$ &
\end{tabular}%
\end{equation}%
for all values of $\mu =1,2,3,4$, or equivalently like $d\frac{\sqrt{5}}{2}%
K_{F}.\lambda _{l}=\frac{2\pi }{5}N+2\pi N_{l}.$ Notice that setting $%
\mathbf{k}=\mathbf{K}_{F}+\mathbf{q}$ with small $q=\left\Vert \mathbf{q}%
\right\Vert $ and expanding $D$ and $\bar{D}$, eq(\ref{DE}) gets reduced to
the following familiar wave equation in Dirac theory
\begin{equation}
\begin{tabular}{llll}
$\frac{d\sqrt{5}}{2}\dsum\limits_{\mu =1}^{4}\mathbf{q}_{\mu }\left(
\begin{array}{cc}
0 & \sigma ^{\mu } \\
\bar{\sigma}^{\mu } & 0%
\end{array}%
\right) \left(
\begin{array}{c}
\phi _{\mathbf{k}} \\
\bar{\chi}_{\mathbf{k}}%
\end{array}%
\right) $ & $=$ & $E\left(
\begin{array}{c}
\phi _{\mathbf{k}} \\
\bar{\chi}_{\mathbf{k}}%
\end{array}%
\right) $ & .%
\end{tabular}%
\end{equation}

\section{Graphene and lattice QCD}

In this section, we would like to deepen the connection between \emph{2D}
graphene and \emph{4D} lattice QCD. This connection has been first noticed
by M.Creutz \textrm{\citep{A3}} and has been developed by several authors
seen its convenience for numerical simulations in QCD .

\subsection{More on link graphene/lattice QCD}

\emph{2D} graphene has some remarkable properties that can be used to
simulate \emph{4D\ }lattice QCD. Besides chirality, one of the interesting
properties is the existence of two Dirac points that can be interpreted as
the light quarks up and down. This follows from the study of the zero modes
of the $2\times 2$ Dirac operator which corresponds also to solve the
vanishing of the following energy dispersion relation%
\begin{equation}
\begin{tabular}{ll}
$\dsum\limits_{l=1}^{l}\cos ak_{l}+i\dsum\limits_{l=1}^{l}\sin ak_{l}=0$ & ,%
\end{tabular}%
\end{equation}%
which has two zeros as given by (\ref{2Z}).\newline
To make contact with lattice QCD, we start by recalling the usual \emph{4D}
hamiltonian density of a free Dirac fermion $\Psi =\left( \psi ^{1},\psi
^{2},\bar{\chi}_{\dot{1}},\bar{\chi}_{\dot{2}}\right) $ living in a
euclidian space time,
\begin{equation}
H=\frac{1}{2}\int d^{4}x\left( \dsum\limits_{\mu =1}^{4}\bar{\Psi}\left(
x\right) \gamma ^{\mu }\frac{\partial \Psi \left( x\right) }{\partial x^{\mu
}}+hc\right) ,  \label{L}
\end{equation}%
where $\mathrm{\gamma }^{\mu }$ are the usual $4\times 4$ Dirac matrices
given by (\ref{515}). Then, we discretize this energy density H by thinking
about the spinorial waves $\Psi \left( x^{1},...,x^{4}\right) $ as $\Psi _{%
\mathbf{r}_{n}}$ living at the $\mathbf{r}_{n}$-nodes of a four dimensional
lattice $\mathbb{L}_{4}$ and its space time gradient $\frac{\partial \Psi
\left( x\right) }{\partial x^{\mu }}$ like $\frac{1}{a}\left( \Psi _{\mathbf{%
r}_{n}+a\mathbf{\mu }}-\Psi _{\mathbf{r}_{n}}\right) $. The field $\Psi _{%
\mathbf{r}_{n}+a\mathbf{\mu }}$ is the value of the Dirac spinor at the
lattice position $\mathbf{r}_{n}+a\mathbf{\mu }$ with the unit vectors $%
\mathbf{\mu }$ giving the four relative positions of the first nearest
neighbors of $\mathbf{r}_{n}$. Putting this discretization back into (\ref{L}%
), we end with the free fermion model
\begin{equation}
\begin{tabular}{ll}
$H=\frac{1}{2a}\dsum\limits_{\mathbf{r}_{n}}\left( \dsum\limits_{\mu =1}^{4}%
\left[ \bar{\Psi}_{\mathbf{r}_{n}}\gamma ^{\mu }\Psi _{\mathbf{r}_{n}+a%
\mathbf{\mu }}-\bar{\Psi}_{\mathbf{r}_{n}+a\mathbf{\mu }}\gamma ^{\mu }\Psi
_{\mathbf{r}_{n}}\right] \right) $ & .%
\end{tabular}
\label{4D}
\end{equation}%
The extra two term $\bar{\Psi}_{\mathbf{r}_{n}}\Gamma \Psi _{\mathbf{r}_{n}}$
and $\left( \bar{\Psi}_{\mathbf{r}_{n}}\Gamma \Psi _{\mathbf{r}_{n}}\right)
^{+}$ with $\Gamma =\frac{1}{2}\sum_{\mu }\gamma ^{\mu }$ cancel each other
because of antisymmetry of the spinors. Clearly, this hamiltonian looks like
the tight binding hamiltonian describing the electronic properties of the
\emph{2D} graphene; so one expects several similarities for the two systems.%
\newline
Mapping the hamiltonian (\ref{4D}) to the Fourier space, we get $H=\sum_{%
\mathbf{k}}\left( \bar{\Psi}_{\mathbf{k}}\mathcal{D}\Psi _{\mathbf{k}%
}\right) $ with Dirac operator $\mathcal{D}=\frac{i}{a}\sum_{\mu
=1}^{4}\gamma ^{\mu }\sin \left( ak_{\mu }\right) ,$where we have set $%
k_{\mu }=\left( \mathbf{k.\mu }\right) $; giving the wave vector component
along the $\mathbf{\mu }$-direction. The $\mathcal{D}$- operator is a $%
4\times 4$ matrix that depends on the wave vector components $\left(
k_{1},k_{2},k_{3},k_{4}\right) $ and has $2^{4}$ zeros located as%
\begin{equation}
\begin{tabular}{llllllll}
$k_{1}=0,\text{ }\frac{\pi }{a}$ & $;$ & $k_{2}=0,\text{ }\frac{\pi }{a}$ & $%
;$ & $k_{3}=0,\text{ }\frac{\pi }{a}$ & $;$ & $k_{4}=0,\text{ }\frac{\pi }{a}
$ & $.$%
\end{tabular}%
\end{equation}%
However, to apply these formalism to 4D lattice QCD, the number of the zero
modes of the Dirac operator should be two in order to interpret them as the
light quarks up and down. Following \textrm{\citep{A3}}, this objective can
be achieved by modifying (\ref{4D}) so that the Dirac operator takes the form%
\begin{equation}
\mathcal{D}=\frac{i}{a}\dsum\limits_{\mu =1}^{4}\gamma ^{\mu }\sin \left(
ak_{\mu }\right) +\frac{i}{a}\dsum\limits_{\mu =1}^{4}\gamma ^{\prime \mu
}\cos \left( ak_{\mu }\right)
\end{equation}%
where $\mathrm{\gamma }^{\prime \mu }$ is some $4\times 4$ matrix that is
introduced in next subsection.

\subsection{\emph{Boriçi-Creutz }fermions}

Following \textrm{\citep{F1,F2}} and using the 4-component Dirac spinors $%
\Psi _{\mathbf{r}}=\left( \phi _{\mathbf{r}}^{a},\bar{\chi}_{\mathbf{r}}^{%
\dot{a}}\right) $, the \emph{Boriçi-Creutz (BC) }lattice action of \emph{free%
} fermions reads in the position space, by dropping mass term $m_{0}$, as
follows:%
\begin{equation}
\begin{tabular}{lll}
$H_{BC}$ $\sim $ & $\frac{1}{2a}\dsum\limits_{\mathbf{r}}\left(
\dsum\limits_{\mu =1}^{4}\bar{\Psi}_{\mathbf{r}}\Upsilon ^{\mu }\Psi _{%
\mathbf{r+}a\mathbf{\mu }}-\dsum\limits_{\mu =1}^{4}\bar{\Psi}_{\mathbf{r+a%
\mathbf{\mu }}}\bar{\Upsilon}^{\mu }\Psi _{\mathbf{r}}\right) -\frac{2i}{a}%
\dsum\limits_{\mathbf{r}}\bar{\Psi}_{\mathbf{r}}\Gamma \Psi _{\mathbf{r}}$ &
\end{tabular}
\label{BBC}
\end{equation}%
where, for simplicity, we have dropped out gauge interactions; and where $%
\Upsilon ^{\mu }=\gamma ^{\mu }+i\gamma ^{\prime \mu }$; which is a kind of
complexification of the Dirac matrices. \newline
Moreover, the matrix $\Gamma $ appearing in the last term is a $4\times 4$
matrix linked to $\gamma ^{\mu }$, $\gamma ^{\prime \mu }$ as follows:
\begin{equation}
\begin{tabular}{lllll}
$\gamma ^{\prime \mu }=\Gamma -\gamma ^{\mu },$ & $2\Gamma
=\dsum\limits_{\mu =1}^{4}\gamma ^{\mu }$, & $\gamma ^{\mu }+i\gamma
^{\prime \mu }=\Upsilon ^{\mu },$ & $\gamma ^{\mu }\gamma ^{\nu }+\gamma
^{\nu }\gamma =2\delta ^{\mu \nu \mu }$ & ,%
\end{tabular}
\label{AG}
\end{equation}%
Mapping (\ref{BBC}) to the reciprocal space, we have
\begin{equation}
\begin{tabular}{lll}
$H_{BC}$ $\sim $ & $\dsum\limits_{\mathbf{k}}\bar{\Psi}_{\mathbf{k}}\mathcal{%
D}_{BC}\Psi _{\mathbf{k}}$ &
\end{tabular}
\label{DC}
\end{equation}%
where the massless Dirac operator $\mathcal{D}_{BC}$ is given by
\begin{equation}
\begin{tabular}{lll}
$\mathcal{D}_{BC}=$ & $+\frac{1}{2a}\left( \Upsilon _{\mu }-\bar{\Upsilon}%
_{\mu }\right) \cos \left( ak_{\mu }\right) $ &  \\
& $+\frac{i}{2a}\left( \Upsilon _{\mu }+\bar{\Upsilon}_{\mu }\right) \sin
\left( ak_{\mu }\right) $ $-\frac{2i}{a}\Gamma $ & .%
\end{tabular}
\label{TIT}
\end{equation}%
Upon using $\Upsilon _{\mu }+\bar{\Upsilon}_{\mu }=2\gamma _{\mu }$ and $%
\Upsilon _{\mu }-\bar{\Upsilon}_{\mu }=2i\gamma _{\mu }^{\prime }$, we can
put $\mathcal{D}_{BC}$ in the form
\begin{equation}
\mathcal{D}_{BC}=D_{\mathbf{k}}+\bar{D}_{\mathbf{k}}-\frac{2i}{a}\Gamma
\label{DIR}
\end{equation}%
with
\begin{equation}
\begin{tabular}{llll}
$D_{\mathbf{k}}=\frac{i}{a}\left( \dsum\limits_{\mu =1}^{4}\gamma ^{\mu
}\sin ak_{\mu }\right) $ & , & $\bar{D}_{\mathbf{k}}=\frac{i}{a}\left(
\dsum\limits_{\mu =1}^{4}\gamma ^{\prime \mu }\cos ak_{\mu }\right) $ & ,%
\end{tabular}
\label{IR}
\end{equation}%
where $k_{\mu }=\mathbf{k}.\mathbf{\mu }$. In the next subsection, we will
derive the explicit expression of these $k_{\mu }$'s in terms of the weight
vectors $\mathbf{\lambda }_{l}$ of the 5-dimensional representation of the $%
SU\left( 5\right) $ symmetry as well as useful relations. \newline
The zero modes of $\mathcal{D}_{BC}$ are points in the reciprocal space;
they are obtained by solving $\mathcal{D}_{BC}=0$; which leads to the
following condition
\begin{equation}
\begin{tabular}{ll}
$\dsum\limits_{\mu =1}^{4}\mathrm{\gamma }^{\mu }\left( \sin aK_{\mu }-\cos
aK_{\mu }\right) -\Gamma \left( 2-\dsum\limits_{\mu =1}^{4}\cos aK_{\mu
}\right) =0$ & .%
\end{tabular}%
\end{equation}%
This condition is a constraint relation on the wave vector components $%
K_{\mu }$; it is solved by the two following wave vectors:%
\begin{equation}
\begin{tabular}{llll}
point $K_{BC}$ & : & $K_{1}=K_{2}=K_{3}=K_{4}=0$ & , \\
point $K_{BC}^{\prime }$ & : & $K_{1}^{\prime }=K_{2}^{\prime
}=K_{3}^{\prime }=K_{4}^{\prime }=\frac{\pi }{2a}$ & ,%
\end{tabular}%
\end{equation}%
that are interpreted in lattice QCD as associated with the light quarks up
and down. \newline
Notice that if giving up the $\gamma _{\mu }^{\prime }$- terms in eqs(\ref%
{BBC}-\ref{DC}); i.e $\gamma _{\mu }^{\prime }\rightarrow 0$, the remaining
terms in $D_{BC}$ namely $D_{\mathbf{k}}\sim \gamma ^{\mu }\sin aK_{\mu }$
have 16 zero modes given by the wave components $K_{\mu }=0,\pi $. By
switching on the $\gamma _{\mu }^{\prime }$-terms, \emph{14} zeros are
removed.

\subsection{Hyperdiamond model}

The hamiltonian $H_{BC}$ is somehow very particular; it let suspecting to
hide a more fundamental property which can be explicitly exhibited by using
hidden symmetries. To that purpose, notice that the price to pay for getting
a Dirac operator with two zero modes is the involvement of the complexified
Dirac matrices $\Upsilon ^{\mu },$ $\bar{\Upsilon}^{\mu }$ as well as the
particular matrix $\Gamma $. Despite that it violates explicitly the $%
SO\left( 4\right) $ Lorentz symmetry since it can be written as
\begin{equation}
\begin{tabular}{ll}
$\Gamma =\frac{1}{2}\dsum\limits_{\mu =1}^{4}\gamma ^{\mu }\upsilon _{\mu },$
& $\upsilon _{\mu }=\left(
\begin{array}{c}
1 \\
1 \\
1 \\
1%
\end{array}%
\right) ,$%
\end{tabular}%
\end{equation}%
the matrix $\Gamma $ plays an important role in studying the zero modes. The
expression of the matrix $\Gamma $ (\ref{AG}) should be thought of as
associated precisely with the solution of the constraint relation $2\Gamma
-\sum_{\mu =1}^{4}\gamma ^{\mu }=0$ that is required by a hidden symmetry of
the \emph{BC} model. This invariance is precisely the $SU\left( 5\right) $
symmetry of the \emph{4D} hyperdiamond to be identified below. Moreover, the
\emph{BC}\ hamiltonian $H_{BC}$ lives on a \emph{4D} lattice $\mathbb{L}%
_{4}^{BC}$ generated by $\mathbf{\mu \equiv v}_{\mu }$; i.e the vectors
\begin{equation}
\begin{tabular}{lllll}
$\mathbf{v}_{1}\mathbf{=}\left(
\begin{array}{c}
\mathbf{v}_{1}^{x} \\
\mathbf{v}_{1}^{y} \\
\mathbf{v}_{1}^{z} \\
\mathbf{v}_{1}^{t}%
\end{array}%
\right) ,$ & $\mathbf{v}_{2}\mathbf{=}\left(
\begin{array}{c}
\mathbf{v}_{2}^{x} \\
\mathbf{v}_{2}^{y} \\
\mathbf{v}_{2}^{z} \\
\mathbf{v}_{2}^{t}%
\end{array}%
\right) ,$ & $\mathbf{v}_{3}\mathbf{=}\left(
\begin{array}{c}
\mathbf{v}_{3}^{x} \\
\mathbf{v}_{3}^{y} \\
\mathbf{v}_{3}^{z} \\
\mathbf{v}_{3}^{t}%
\end{array}%
\right) ,$ & $\mathbf{v}_{4}\mathbf{=}\left(
\begin{array}{c}
\mathbf{v}_{4}^{x} \\
\mathbf{v}_{4}^{y} \\
\mathbf{v}_{4}^{z} \\
\mathbf{v}_{4}^{t}%
\end{array}%
\right) $ &
\end{tabular}%
\end{equation}%
These\ $\mathbf{\mu }$-vectors look somehow ambiguous to be interpreted both
by using the analogy with \emph{4D} graphene prototype; and also from the $%
SU\left( 5\right) $ symmetry view.\ Indeed, to each site $\mathbf{r}\in
\mathbb{L}_{4}^{BC}$ there should be \emph{5} first nearest neighbors that
are rotated by $SU\left( 5\right) $ symmetry. But from the \emph{BC}
hamiltonian we learn that the first nearest neighbors to each site $\mathbf{r%
}$ are:%
\begin{equation}
\begin{tabular}{lll}
$\mathbf{r}$ & $\rightarrow $ & $\left\{
\begin{array}{c}
\mathbf{r+}a\mathbf{v}_{1} \\
\mathbf{r+}a\mathbf{v}_{2} \\
\mathbf{r+}a\mathbf{v}_{3} \\
\mathbf{r+}a\mathbf{v}_{4}%
\end{array}%
\right. $.%
\end{tabular}%
\end{equation}%
The fifth missing one, namely $\mathbf{r}\rightarrow \mathbf{r+}a\mathbf{v}%
_{5}$ may be interpreted in the \emph{BC} fermions as associated with the
extra term involving the matrix $\Gamma $. To take into account the five
nearest neighbors, we have to use the rigorous correspondence $\Gamma ^{\mu
}\rightarrow \mathbf{v}_{\mu }$ and $\Gamma ^{5}\rightarrow \mathbf{v}_{5}$
which can be also written in a combined form as follows $\Gamma
^{M}\rightarrow \mathbf{v}_{M}$ with $\Gamma ^{M}=\left( \Gamma ^{\mu
},\Gamma ^{5}\right) $ and $\mathbf{v}_{M}=\left( \mathbf{v}_{\mu },\mathbf{v%
}_{5}\right) .$ Because of the $SU\left( 5\right) $ symmetry properties, we
also have to require the condition $\mathbf{v}_{1}+\mathbf{\mathbf{v}_{2}+%
\mathbf{v}_{3}+\mathbf{v}_{4}+v}_{5}=0$ characterizing the \emph{5} first
nearest neighbors. To determine the explicit expressions of the matrices $%
\Gamma _{M}$ in terms of the usual Dirac ones, we modify the \emph{BC} model
(\ref{BBC}) as follows%
\begin{equation}
\begin{tabular}{lll}
$H_{BC}^{\prime }$ $\sim $ & $\frac{1}{2a}\dsum\limits_{\mathbf{r}}\left(
\dsum\limits_{M=1}^{5}\bar{\Psi}_{\mathbf{r}}\Gamma ^{M}\Psi _{\mathbf{r+}a%
\mathbf{v}_{M}}-\dsum\limits_{M=1}^{5}\bar{\Psi}_{\mathbf{r+av}_{M}}\Gamma
^{M}\Psi _{\mathbf{r}}\right) $ & ,%
\end{tabular}
\label{EXT}
\end{equation}%
exhibiting both $SO\left( 4\right) $ and $SU\left( 5\right) $ symmetries and
leading to the following free Dirac operator%
\begin{equation}
\begin{tabular}{lll}
$\mathcal{D}=$ & $\frac{i}{2a}\dsum\limits_{\mu =1}^{4}\left( \Gamma _{\mu }+%
\bar{\Gamma}_{\mu }\right) \sin \left( ak_{\mu }\right) +\frac{i}{2a}\left(
\Gamma _{5}+\bar{\Gamma}_{5}\right) \sin \left( ak_{5}\right) $ &  \\
& $\frac{1}{2a}\dsum\limits_{\mu =1}^{4}\left( \Gamma _{\mu }-\bar{\Gamma}%
_{\mu }\right) \cos \left( ak_{\mu }\right) +\frac{1}{2a}\left( \Gamma _{5}-%
\bar{\Gamma}_{5}\right) \cos \left( ak_{5}\right) $ &
\end{tabular}
\label{EG}
\end{equation}%
where $k_{M}=\mathbf{k.v}_{M}$ and where $\Pi _{M=1}^{5}\left(
e^{iak_{M}}\right) =1$, $\sum_{M=1}^{5}k_{M}=0$ expressing the conservation
of total momenta at each lattice site. Equating with (\ref{TIT}-\ref{DIR}-%
\ref{IR}), we get the identities%
\begin{equation}
\begin{tabular}{ll}
$\Upsilon _{\mu }+\bar{\Upsilon}_{\mu }=\Gamma _{\mu }+\bar{\Gamma}_{\mu },$
& $\Upsilon _{\mu }-\bar{\Upsilon}_{\mu }=\Gamma _{\mu }-\bar{\Gamma}_{\mu
}, $%
\end{tabular}
\label{Ga}
\end{equation}%
and
\begin{equation}
\begin{tabular}{ll}
$\frac{i}{2a}\left( \Gamma _{5}+\bar{\Gamma}_{5}\right) \sin \left(
ak_{5}\right) +\frac{1}{2a}\left( \Gamma _{5}-\bar{\Gamma}_{5}\right) \cos
\left( ak_{5}\right) =-\frac{4i}{2a}\Gamma $ & .%
\end{tabular}%
\end{equation}%
Eqs(\ref{Ga}) are solved by $\Gamma _{\mu }=\Upsilon _{\mu }$; that is $%
\Gamma _{\mu }=\gamma ^{\mu }+i\left( \Gamma -\gamma ^{\mu }\right) $ while
\begin{equation}
\begin{tabular}{llll}
$\Gamma _{5}=-2i\Gamma $ & for & $\sin \left( ak_{5}\right) =0$ & , \\
$\Gamma _{5}=-2\Gamma $ & for & $\sin \left( ak_{5}\right) =1$ & .%
\end{tabular}%
\end{equation}%
where $k_{5}=-\left( k_{1}+k_{2}+k_{3}+k_{4}\right) $. In this 5-dimensional
approach, the ambiguity in dealing with the $\mathbf{\mu }$-vectors is
overcome; and the underlying $SO\left( 4\right) $ and $SU\left( 5\right) $
symmetries of the model in reciprocal space are explicitly exhibited.

\section{Conclusion and comments}

Being a simple lattice-carbon based structure with delocalized electrons,
graphene has been shown to exhibit several exotic physical properties and
chemical reactions leading to the synthesis of graphene type derivatives
such as graphAne and graphOne. In this book chapter, we have shown that
graphene has also very remarkable hidden symmetries that capture basic
physical properties; one of these symmetries is the well known $SU\left(
2\right) $ invariance of the unit cells that plays a crucial role in the
study of the electronic properties using first principle calculations.
Another remarkable hidden invariance, which has been developed in this work,
is the \emph{SU}$\left( 3\right) $ symmetry that captures both
crystallographic and physical properties of the graphene. For instance,
first nearest neighbors form 3-dimensional representations of $SU\left(
3\right) $; and the second nearest neighbor ones transform in its adjoint.
Moreover, basic constraint relations like $\mathbf{\upsilon }_{1}+\mathbf{%
\upsilon }_{2}+\mathbf{\upsilon }_{3}=0$ is precisely a $SU\left( 3\right) $
group property; and its solutions are exactly given by group theory.
Furthermore, the location of the Dirac zero modes of graphene is also
captured by $SU\left( 3\right) $ seen that these points are given by $\pm
\frac{2\pi }{3d}\mathbf{\alpha }_{1},$ $\pm \frac{2\pi }{3d}\mathbf{\alpha }%
_{2},$ $\pm \frac{2\pi }{3d}\mathbf{\alpha }_{3}$ where the $\mathbf{\alpha }%
_{i}$'s are the $SU\left( 3\right) $ roots that generate the reciprocal
space. \newline
On the other hand, from $SU\left( 3\right) $ group theory's point of view,
graphene has cousin systems with generic $SU\left( N\right) $\ symmetries
where the integer $N$ takes the values $2,3,4,...$. The leading graphene
cousin systems are linear molecules with hidden SU$\left( 2\right) $
invariance; this is precisely the case of poly-acetylene, cumulene and
poly-yne studied in section 4. The graphene cousin systems with hidden SU$%
\left( 4\right) $ and SU$\left( 5\right) $ symmetries are given by \emph{3D}
diamond; and \emph{4D} hyperdiamond which has an application in \emph{4D}-
lattice QCD. \newline
Finally, it is worth to mention that the peculiar and unique properties of
graphene are expected to open new areas of applications due to its important
electronic, spintronic, mechanical and optical properties. The challenge is
find low-cost-processes for producing graphene and graphene-based structures
and to tune its properties to the targeted applications such as the
replacement of silicon in the field of new-type of semiconductors and new
electronics, new data-storage devices, new materials with exceptional
mechanical properties and so on.\newline
Various attempts are also made to incorporate other atoms within the
structure of graphene or combine the graphene-based structures with other
materials in sandwich type structure or in chemical way by binding it to
various molecules with divers topologies and functionalities.

\end{document}